\documentclass[useAMS,usenatbib]{mn2e}
\usepackage{graphicx,amssymb,dcolumn,color}
\newcommand{\figref}[1]{figure~\ref{#1}}
\newcommand{\Figref}[1]{Figure~\ref{#1}}
\newcommand{\secref}[1]{section~\ref{#1}}
\newcommand{\secsand}[2]{sections~\ref{#1} and \ref{#2}}

\newcommand{\apref}[1]{appendix~\ref{#1}}

\newcommand{\Eqref}[1]{Equation~(\ref{#1})}
\newcommand{\eqref}[1]{equation~(\ref{#1})}

\newcommand{\Eqsand}[2]{Equations~(\ref{#1}) and (\ref{#2})}
\newcommand{\eqsand}[2]{equations~(\ref{#1}) and (\ref{#2})}

\newcommand{\eqsdash}[2]{equations~(\ref{#1}--\ref{#2})}
\newcommand{\exref}[1]{(\ref{#1})}

\newcommand{\bea}{\begin{eqnarray}}
\newcommand{\eea}{\end{eqnarray}}
\newcommand{\beq}{\begin{equation}}
\newcommand{\eeq}{\end{equation}}
\newcommand{\lt}{\left}
\newcommand{\rt}{\right}
\newcommand{\bl}{\bigl}
\newcommand{\br}{\bigr}
\renewcommand{\la}{\langle}
\newcommand{\ra}{\rangle}
\newcommand{\dd}{\partial}
\newcommand{\const}{{\rm const}}
\newcommand{\eps}{\varepsilon}
\newcommand{\rmd}{\rmn{d}}
\renewcommand{\Re}{\rmn{Re}}

\newcommand{\vthi}{v_{\mathrm{th}i}}

\newcommand{\nuii}{\nu_{ii}} 
\newcommand{\nueff}{\nu_{\rm eff}} 
\newcommand{\mueff}{\mu_{\rm eff}} 
\newcommand{\tsigma}{\tilde\sigma} 
\newcommand{\tgam}{\bar\gamma} 
\newcommand{\gmir}{\gamma_{\rm m}} 
\newcommand{\gfir}{\gamma_{\rm f}} 
\newcommand{\lmir}{l_{\rm m}} 
\newcommand{\lfir}{l_{\rm f}} 
\renewcommand{\tt}{\tau} 
\newcommand{\Dtt}{\Delta\tt} 
\newcommand{\tdyn}{t_{\rm dyn}} 
\newcommand{\ttdyn}{\tt_{\rm dyn}} 
\newcommand{\fmax}{f_{\rm max}} 

\newcommand{\tP}{\tilde P} 
\newcommand{\tC}{\hat C} 
\newcommand{\LL}{\hat L}

\newcommand{\pperp}{p_\perp}
\newcommand{\ppar}{p_\parallel}

\newcommand{\vB}{\bmath{B}}

\newcommand{\vb}{\hat{\bmath{b}}}

\newcommand{\vu}{\bmath{u}}

\newcommand{\vperp}{v_\perp}

\newcommand{\vdel}{\bmath{\nabla}}

\title[Magnetic-field evolution in extragalactic plasmas]{Models of magnetic-field evolution 
and effective viscosity in weakly collisional extragalactic plasmas}
\author[F.~Mogavero and A. A. Schekochihin]{
Federico~Mogavero$^{1,2,3}$
and
Alexander~A.~Schekochihin$^{2,3}$\thanks{E-mail: a.schekochihin1@physics.ox.ac.uk}\\
$^{1}$ICFP, D\'epartement de Physique, 
\'Ecole Normale Sup\'erieure, 24 rue Lhomond, 75252 Paris Cedex 05, France\\
$^{2}$The Rudolf Peierls Centre for Theoretical Physics, University of Oxford, 
1 Keble Rd, Oxford OX1 3NP, U.K.\\
$^{3}$Merton College, Merton St, Oxford OX1 4JD, U.K.}

\begin{document}

\date{\today}

\pagerange{\pageref{firstpage}--\pageref{lastpage}} \pubyear{2013}

\maketitle

\label{firstpage}

\begin{abstract}

In weakly collisional plasmas such as the intracluster 
medium (ICM), the viscous stress and the rate of change of the 
magnetic energy are proportional to the local pressure anisotropy, 
so subject to constraints imposed by the pressure-anisotropy-driven 
microinstabilities (mirror and firehose) and controlled by the local instantaneous 
plasma $\beta$. The dynamics of such plasmas 
can be dramatically different from a conventional MHD fluid. 
The plasma is expected to stay locally marginal with respect to the 
instabilities, but how it does this remains an open question. 
Two models of magnetic-field evolution are investigated. In the first, marginality is 
achieved via suppression of the rate of change of the field.  
In the second, the instabilities give rise to anomalous collisionality, 
reducing pressure anisotropy to marginal --- at the same time decreasing viscosity and 
so increasing the turbulent rate of strain.
Implications of these two models are studied in a simplified zero-dimensional 
setting. In the first model, the field grows explosively but on a time scale that 
scales with the initial $\beta$, while in the second, dynamical field strength can be reached in one large-scale turbulence 
turn-over time regardless of the initial seed. Both models produce very 
intermittent fields.  
Both also suffer from fairly strong constraints on their applicability: 
for typical cluster-core conditions, 
scale separation between the fluid motions (with account of 
suppressed viscous stress) and the miscoscale fluctuations 
breaks down at $\beta\sim10^{4}-10^{5}$. 
At larger $\beta$ (weaker fields), a fully collisionless 
plasma dynamo theory is needed to justify 
field growth from a tiny primordial seed. However, the models discussed here are 
appropriate for studying the structure of the 
currently observed field as well as large-scale dynamics and thermodynamics of the magnetized ICM 
or similarly dilute astrophysical plasmas.

\end{abstract}


\section{Introduction}

Both analytical theory and numerical modelling of the large-scale dynamics of extragalactic plasmas 
present conceptual challenges that are more serious than merely constraints of numerical resolution 
or analytical tractability. One of the most intriguing of these challenges is understanding what 
happens in a weakly collisional plasma when a dynamically small magnetic field is stretched 
and tangled by the plasma flows --- a process that is both interesting in itself, in the context 
of the genesis of the magnetic fields ubiquitously observed in the Universe, and integral to 
any large-scale fluid dynamics of astrophysical plasmas. 

We call a plasma {\em weakly collisional} and {\em magnetized} 
when, on the one hand, the collision frequency in it exceeds the typical frequencies associated 
with fluid motions,
waves or instabilities, but, on the other hand, it is much smaller than the Larmor frequency 
of the plasma's constituent ions and electrons gyrating around the magnetic 
field (\citealt{Balbus04} calls such plasmas {\em dilute}). While this regime requires 
some magnetic field to be present, this field by no means needs to be dynamically significant. 
As an example, consiser the intracluster medium (ICM) in the cores of galaxy 
clusters. The ratio of the ion collision frequency $\nuii$ to the ion Larmor frequency 
$\Omega_i$, when referred to the conditions typical of this environment 
\citep[taken from][]{Rosin11} turns out to be
\beq
\frac{\nuii}{\Omega_i} \sim \lt(\frac{B}{10^{-17}~{\rm G}}\rt)^{-1}, 
\eeq 
where $B$ is the magnetic-field strength.
Thus, $B\sim 10^{-17}$~G (corresponding to the ratio of the 
thermal to magnetic energy $\beta=8\pi p/B^2 \sim 10^{24}$)  
is sufficient for the plasma to be magnetized, but magnetic field is 
not strong enough to affect plasma motions via the Lorentz force until 
$B\sim 10^{-6}$~G ($\beta\sim 10^2$), a value at which the magnetic-energy density 
is comparable to the kinetic-energy density of 
the turbulent plasma flows at the viscous scale \citep{Sch06}. 
Note that this is close to the values of $B$ measured 
in cluster cores \citep[e.g.,][]{Carilli02,Govoni04,Vogt05}.  

In a weakly collisional magnetized plasma, the magnetic moment of each particle, 
$\mu = \vperp^2/B$ (where $\vperp$ is the peculiar velocity of the particle's Larmor motion), 
is conserved on the timescales shorter than the collision time. Therefore, as the magnetic field, 
which is frozen into the turbulent flow and, being dynamically weak, might appear to be entirely at 
its mercy, is stretched and tangled, its strength changes (the field gets larger in some places, 
weaker in others) and pressure anisotropies develop. If we ignore heat fluxes 
and assume that flows are subsonic (therefore, incompressible), the local pressure 
anisotropy is\footnote{This follows from the so-called CGL equations \citep{CGL}, 
with collisions retained \citep[see, e.g.,][]{Sch10}. Besides $\mu$ conservation, 
these equations are also an expression of the conservation of the so-called 
longitudinal invariant \citep[e.g.,][]{Kulsrud83,Quest96}, which is related to the bounce 
invariant of particles trapped in the local inhomogeneities of the magnetic-field 
strength.}
\beq
\Delta \equiv \frac{\pperp-\ppar}{p} \approx \frac{1}{\nuii}\frac{1}{B}\frac{\rmd B}{\rmd t},
\label{eq:Delta}
\eeq
where $\pperp$ ($\ppar$) is the perpendicular (parallel) pressure 
with respect to the local direction of the field. 
The rate of change of $B$ can be related to the local plasma flow 
velocity $\vu$ according to 
\beq
\frac{\rmd B}{\rmd t} = (\vb\vb:\vdel\vu) B \equiv \gamma B
\label{eq:B}
\eeq
(this follows from the MHD induction equation). 
Here $\rmd/\rmd t = \dd_t + \vu\cdot\vdel$ and $\vb = \vB/B$, 
so $\gamma=\vb\vb:\vdel\vu$ is the local field-stretching rate. 
All fluid frequencies, $\gamma$ amongst them, are taken to be small compared to $\nuii$, 
so $\Delta\ll1$. For typical core ICM parameters, $\Delta\sim0.01$ \citep{Rosin11,Kunz11}.  

Even though $\Delta$ is small, it turns out to be sufficient to render the ICM violently 
unstable to the firehose and mirror instabilities \citep[see][and references therein]{Sch05}, 
which have growth rates closer to $\Omega_i$ than to $\nuii$ 
(see \citealt{Davidson68,Yoon93,Hellinger00,Hellinger07,Sch10} and \secref{sec:M1_constraints}) 
and thus can be viewed as instantaneous from the point of view of all collisional and fluid processes 
(i.e., all macroscale dynamics). The instabilities are quenched when 
\beq
\Delta = \frac{\gamma}{\nuii} \in \lt[-\frac{2}{\beta},\frac{1}{\beta}\rt], 
\label{eq:interval}
\eeq
where the lower threshold is for the firehose and the 
upper threshold for the mirror.\footnote{For the mirror, this is only an approximate bound 
assuming cold electrons and bi-Maxwellian ions \citep{Hellinger07} --- approximations that are 
generally incorrect quantitatively but give a threshold that is simple enough and close enough 
to the truth to be useful in a qualitative discussion.} 
Thus, at large enough $\beta$, any change in the magnetic field leads 
to instabilities and in order to understand whether and how magnetic field can continue 
changing, we must account for the effect these instabilities have on the dynamics of the ICM. 

It is not currently clear precisely how these instabilities saturate, but it is clear 
that the result of their saturation will be that the pressure anisotropy averaged 
over scales smaller than those of the fluid motions will 
not stray beyond the marginal-stability boundaries \exref{eq:interval} 
(the clearest evidence for this is found in the solar wind; see \citealt{Kasper02,Hellinger06,Bale09}). 
This suggests that large-scale dynamics of the ICM might perhaps be modelled in total 
ignorance of the microphysical complexities associated with the firehose and mirror 
saturation, simply by assuming that $\Delta$ stays at most marginal, as per \eqref{eq:interval}. 
There are, unfortunately, two very different ways in which this can be accomplished 
and which of them is correct depends on how the instabilities saturate. 

In \eqref{eq:interval}, by $\Delta$ let us understand the {\em mean} pressure anisotropy averaged 
over the microscales at which the unstable fluctuations appear; similarly, 
$\gamma = \overline{\vb\vb:\vdel\vu}$ is the {\em mean} field-stretching rate. 
Whenever $\Delta$ attempts to cross either of the boundaries \exref{eq:interval}, it can be reined 
in {\em either} via $\gamma$ being effectively suppressed 
by the instabilities ({\bf Model~I}) {\em or} via the effective collision rate $\nuii$ being enhanced 
by, say, anomalous particle scattering off the firehose or mirror fluctuations ({\bf Model~II}).  
The two models amount to two very different closures for the fluid equations 
and lead to very different physical consequences (see below).

After providing, in the remainder of this Introduction (\secref{sec:status}), 
some pointers to relevant previous 
literature pertaining to the justification and/or consequences of these two models,  
we will, in the rest of this paper, study, using very drastically simplified equations for 
the evolution of the magnetic field and the local rate of its stretching, what implications 
they might have for the field growth and its spatial distribution. 
A ``zero-dimensional-dynamo'' paradigm that will be the basis for our investigation 
of the two models will be introduced in \secref{sec:zeroD}. 
Model~I will be studied in \secref{sec:M1}, 
Model~II in \secref{sec:M2}. 
The results for each of these will be summarized at the end of the section devoted to it 
and a more global discussion given in \secref{sec:disc}. 

\subsection{Current Status of the Two Models}
\label{sec:status}

The suppressed-stretching-rate model (Model~I) was used 
by \citet{Kunz11} in a theory of ICM thermal stability 
(this will be further discussed in \secref{sec:disc}). 
Various versions of the anomalous-scattering model (Model~II) 
have been used in the theory of explosive dynamo 
\citep{Sch06}, simulations of accretion flows \citep{Sharma06,Sharma07},  
anisotropic heat-conduction instabilities \citep{Kunz12} and of 
turbulent dynamo \citep{Lima14} 
as well as a number of simulations and models of space plasmas 
\citep{Samsonov01,Samsonov07,Chandran11,Meng12}.  
In \secref{sec:modelling}, we will explain why, in light of 
the analysis of \secref{sec:M2}, adequate numerical implementation 
of Model~II may be harder than it appears. 

We would like to avoid a lengthy discussion of the relative microphysical 
merits of the two models but it is perhaps worthwhile to outline the state of play
as we see it. 
Traditionally, modelling results and observational evidence (in the solar wind)
have been interpreted in terms of anomalous scattering 
of particles pinning the pressure anisotropy at the instability 
thresholds \citep[e.g.,][]{Gary97,Gary98,Gary00,Bale09}.  
Some doubt is, however, cast on the possibility of enhanced scattering 
at the mirror threshold by the fact that near marginal stability, the mirror fluctuations 
have scales much larger than the Larmor radius and growth rates much smaller than the 
Larmor frequency \citep{Davidson68,Hellinger07}
and so should not be able to break the conservation of the first adiabatic invariant. 
Similar reasoning (fluctuation scales too large to break adiabatic invariance) 
holds for parallel firehose fluctuations --- but the oblique firehose does 
produce Larmor-scale fluctations \citep{Yoon93,Hellinger00}. 

A weakly nonlinear theory of the parallel firehose \citep{Sch08,Rosin11} 
finds a mechanism for pinning $\Delta$ at marginal stability that relies on effective cancellation 
of the mean rate of change of the magnetic-field strength rather than on anomalous scattering 
(see \secref{sec:M1closure}). An example of magnetic-field evolution in a Braginskii plasma where 
marginality with respect to the firehose is maintained via modification 
of the flow field was recently found by \citet{Melville14}. 
A no-scattering mechanism for maintaining marginal $\Delta$ has also been proposed for the mirror, 
with a special role assigned to trapped particles in setting up magnetic 
``holes'' that compensate for the mean-field growth  
(\citealt{Pantellini95,Kivelson96,Pantellini98,Sch08,Rincon14}; cf.\ \citealt{Califano08,Pokhotelov10}). 

Recent PIC simulations of a shearing, high-$\beta$ plasma by \citet{Kunz14} 
appear to support this line of thinking for the mirror threshold 
(no scattering, nonlinear evolution broadly in agreement with 
\citealt{Sch08,Rincon14}) --- except possibly in cases of very strong 
drive (see also \citealt{Riquelme14}) or at late stages of the nonlinear evolution, 
which may or may not be relevant to a situation where the turbulent velocity field that 
drives the pressure anisotropy decorrelates every turnover time. 
Incidentally, both \citet{Kunz14} and \citet{Riquelme14} confirm that it is the mirror 
instability that dominates in high-$\beta$ regimes with positive pressure anisotropy, 
not the ion-cyclotron instability \citep[cf.][]{Gary97,Gary00}. 
Note that solar-wind measurements also appear to support 
mirror over the ion-cyclotron instability as controlling the positive-pressure-anisotropy 
threshold \citep{Hellinger06}. 

At the firehose threshold, \citet{Kunz14} find evidence both of transient 
evolution reminiscent of the no-scattering theory \citep{Sch08,Rosin11} 
and of oblique fluctuations at Larmor scales vigorously breaking 
adibatic invariance in their saturated nonlinear state. 

Thus, the state of play appears to be in flux and, with the complete theory 
or complete understanding lacking, we take the view that neither of our two models 
can as yet be ruled out --- and that possible implications of either are worth 
investigating.    

\section{Zero-Dimensional Dynamo}
\label{sec:zeroD}

We model the magnetic-field evolution by a ``zero-dimensional'' equation, 
\beq
\dd_t B = \gamma B,
\label{eq:B0}
\eeq
where $\gamma(t)$ is a random time-dependent stretching rate and we will interpret 
different realizations of $\gamma$ as corresponding to different 
(strictly speaking, Lagrangian) spatial positions. 
In the absence of either dynamical back reaction by the field on the flow 
or of any plasma microphysical effects, the stretching rate is of order of 
the local rate of strain in the turbulent flow advecting the field.  
So we set $\gamma=\sigma(t)$, where $\sigma$ is a scalar quantity representing 
this local rate of strain and modelled as an Ornstein-Uhlenbeck process:
\beq
\dd_t\sigma = -2\sigma_0\,\sigma + 2\sigma_0^{3/2}\chi(t),
\label{eq:sigma0}
\eeq
where $\chi(t)$ is a unit Gaussian white noise, $\la\chi(t)\chi(t')\ra = \delta(t-t')$,
and $\sigma_0$ is the rms value of $\sigma$ (in statistically steady state), which is  
also the decorrelation rate (see \apref{ap:zeroD}; the factors of 2 are for 
future normalization covenience). In Kolmogorov turbulence, the largest rate of strain 
is associated with the motions {\em at the viscous scale} (because 
smaller-scale ``eddies'' have shorter turn-over times) and so 
$\sigma_0\sim (\eps/\mu)^{1/2}$, where $\eps$ is the mean power injected into 
the turbulence (originating from large-scale driving mechanisms; in clusters, 
merger-excited instabilities, AGN ejecta, or galaxy wakes; 
see, e.g., \citealt{Norman99,Subramanian06,Ensslin06,Ruszkowski10}) 
and $\mu\sim p/\nuii$ is the dynamical viscosity of the ICM.\footnote{In a magnetized plasma, 
this is the parallel \citet{Braginskii65} viscosity. 
While only the parallel component of the rate-of-strain tensor, 
$\vb\vb:\vdel\vu$, is damped by this viscosity and the perendicular viscosity is 
much smaller, the parallel viscosity is the relevant one for the magnetic-field-stregth 
evolution because only motions with $\vb\vb:\vdel\vu\neq0$ can change $B$.}
Thus, the first term on the right-hand side of \eqref{eq:sigma0} stands for 
viscous damping of the velocity and the second for the fresh input of turbulent 
power into the viscous-scale motions coming from the inertial range. 

In the absence of further constraints, \eqsand{eq:B0}{eq:sigma0} lead to an ensemble 
of realizations of $B$ with exponentially growing moments, 
$\la B^n\ra\propto\exp(\sigma_0 n^2t/2)$, and a lognormal probability 
density, $P(B) = B^{-1}\exp\lt[-(\ln B)^2/2\sigma_0 t\rt]/\sqrt{2\pi\sigma_0 t}$ 
(see \apref{ap:zeroD}). 
This is similar to the standard properties of a kinematic dynamo in a one-scale 
stochastic velocity field \citep[see][and references therein]{Sch02folding,Sch04}. 
Thus, the magnetic energy grows exponentially but the field is quite intermittent
(note that, unlike the magnetic energy, a typical realization of the field 
grows subexponentially, $\ln B\sim \sqrt{\sigma_0 t}$, so only a small 
fraction of the realizations contribute to the exponential growth of $\la B^2\ra$). 

The field will grow until the Lorentz force is strong enough 
to affect the rate of strain. This happens when the magnetic-energy density 
becomes comparable to the kinetic-energy density of the motions at the 
viscous scale:
\beq
\frac{B^2}{8\pi} \sim (\eps\mu)^{1/2} \sim \frac{\sigma_0 p}{\nuii}. 
\label{eq:sat}
\eeq
Since magnetic energy grows exponentially at the rate $\gamma=\sigma\sim\sigma_0$, 
the dynamical strength \exref{eq:sat} is achieved after the time
\beq
\tdyn \sim \frac{1}{\sigma_0}\,\ln\lt(\frac{\sigma_0\beta_0}{\nuii}\rt),
\label{eq:t0}
\eeq
where $\beta_0=8\pi p/B_0^2$ is plasma beta associated with the 
initial (seed) field. 

Further growth of the field does occur after that, but 
requires modelling of its dynamical effect on the flow 
\citep[see][]{Sch02,Sch04,Cho09,Beresnyak12}. The field can typically grow in this 
nonlinear regime by a factor of $\Re^{1/2}$, 
where $\Re = \rho U L/\mu$ is the Reynolds number of the intracluster turbulence
($\rho$ is the mass density and $U$ the typical velocity at the outer scale $L$). 
With the viscosity based on the Coulomb collisionality $\nuii$, 
$\Re$ is not very much larger than unity in the ICM \citep{Sch06,Rosin11,Kunz11}, 
so the difference between the field given by \eqref{eq:sat} and the final 
saturated level is not very large (we will come back to these nonlinear 
issues in \secsand{sec:multiscale}{sec:M2qual}). Observationally, the 
$B\sim1-10\,\mu$G fields found ubiquitously in clusters \citep{Carilli02,Govoni04,Vogt05}
are quite close to the magnitude given by \eqref{eq:sat}; this is also the field magnitude 
that gives marginal values of the pressure anisotropy (see \eqref{eq:interval}). 

The ``zero-dimensional-dynamo'' paradigm is, of course, a gross simplification, 
not least because it says nothing of the spatial structure of the field or of its 
direction relative to the flow and also ignores the effect of resistivity (or whatever other 
flux unfreezing mechanism turns out to be important in a weakly collisional plasma, 
another area of very poor current knowledge). However, it is a useful tool to explore 
what possible effect our two microphysical closure models might have on the evolution 
of the magnetic field. 

\begin{figure*}
\includegraphics[width=85mm]{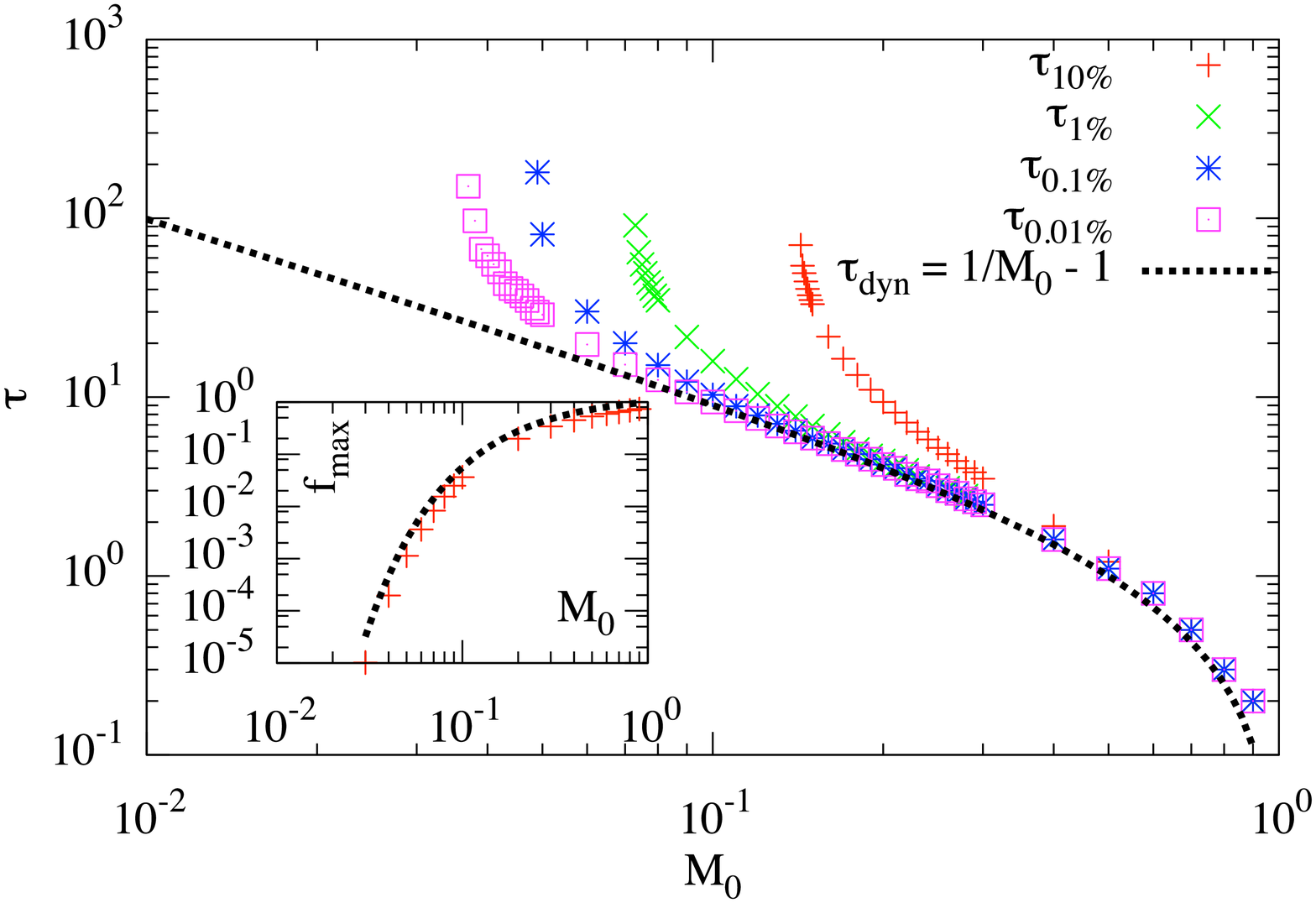}
\includegraphics[width=85mm]{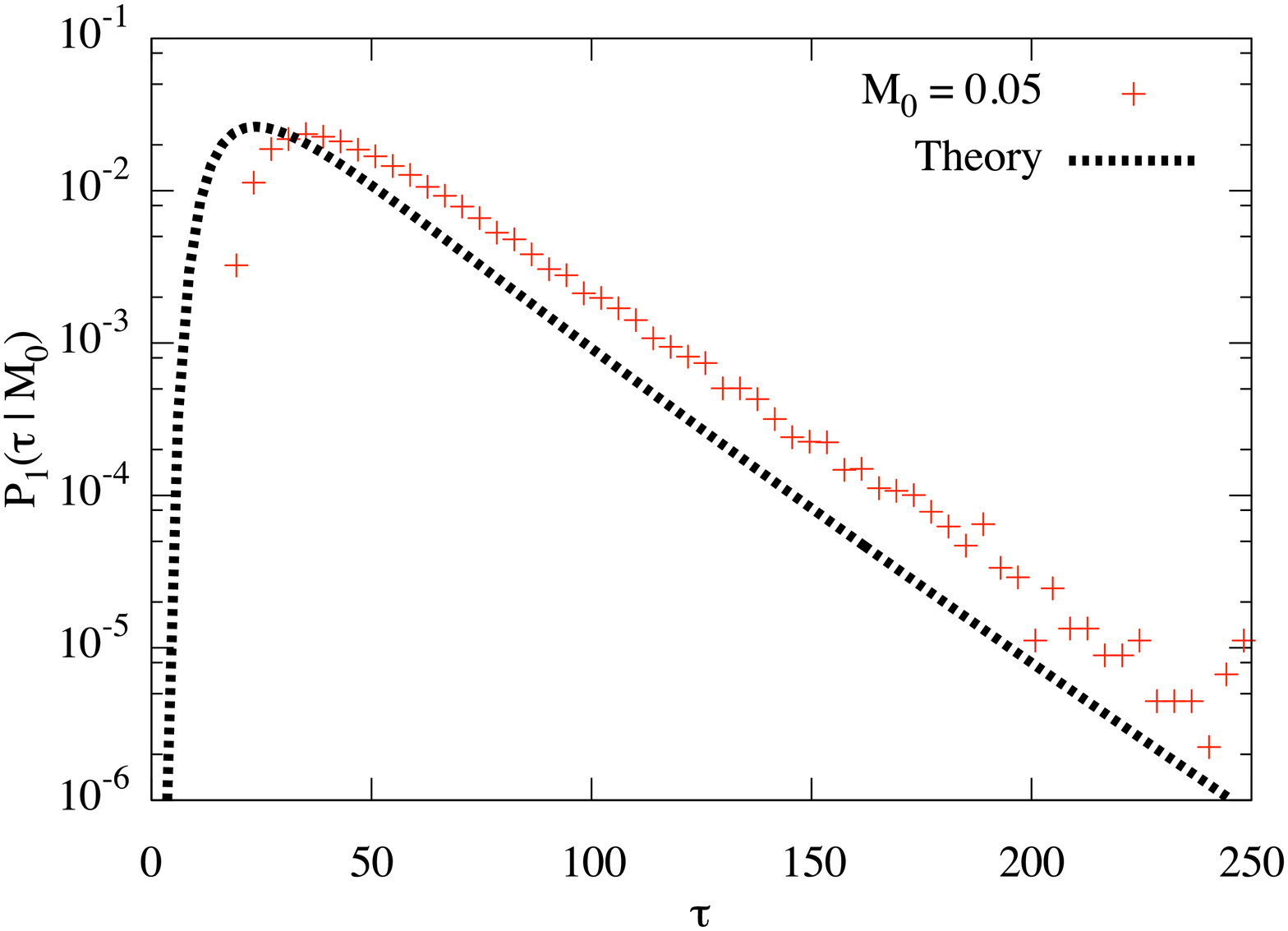}
\caption{{\it Left panel:} Times $\tau_f$ for the fractions $f=10\%$, 1\%, 0.1\%, and 0.01\% 
of all realizations to achieve dynamical field strength 
($M=1$; \eqref{eq:sat}) in Model~I, \eqsand{eq:M1}{eq:sigma1}, 
plotted vs.\ initial normalized field energy $M_0=\nuii/\sigma_0\beta_0$. 
The dotted line is $\ttdyn = 1/M_0 - 1$ (\eqref{eq:t1}). 
{\it Inset:} Total fraction of realizations that ever make it to $M=1$. 
The dotted line is the theoretical curve \exref{eq:fmax_M1}. 
{\it Right panel:} PDF of the time $\tt$ it took those realizations to 
grow to $M=1$. The case of $M_0=0.05$ is shown together with the theoretical 
curve \exref{eq:P1_M1}.} 
\label{fig:M1_tau}
\end{figure*}

\section{Model I: Suppressed stretching rate}
\label{sec:M1}

\subsection{Microphysical Closure} 
\label{sec:M1closure}

In this approach, we assume that the effect of pressure-anisotropy-driven 
instabilities is to suppress the stretching rate $\gamma$ in \eqref{eq:B0} 
whenever the rate of strain $\sigma$ 
becomes large enough to violate the mirror or firehose stability conditions. 
In order to model this, we stipulate that in \eqref{eq:B0}, 
\bea
\label{eq:M1_mirror}
{\rm if}\ \sigma\ge0,&& \gamma = \min\lt\{\sigma,\frac{\nuii}{\beta}\rt\},\\ 
{\rm if}\ \sigma < 0,&& \gamma = \max\lt\{\sigma,-\frac{2\nuii}{\beta}\rt\}, 
\label{eq:M1_fh} 
\eea
i.e., while the collision rate $\nuii$ is always the Coulomb collision rate
and the rate of strain $\sigma$ ranges freely, 
the pressure anisotropy $\Delta$ and, therefore, 
the field-stretching rate $\gamma$ (see \eqref{eq:interval})
cannot cross the stability boundaries. 

An example of how such a closure might be achieved 
microphysically is provided by the calculation by \citet{Rosin11} of the nonlinear 
evolution of the parallel firehose instability (the only analytical example we are aware 
of from which any inferences about an effective closure can be drawn). 
In that calculation, both the large-scale magnetic field and the large-scale 
velocity field are perturbed by fast-growing (and oscillating), small-scale 
firehose fluctuations: $\vB = \vB_0 + \delta\vB_\perp$, $\vu = \vu_0 + \delta\vu_\perp$. 
The mean rate of change of the magnetic field is 
\bea
\nonumber
\frac{1}{\overline{B}}\overline{\frac{\rmd B}{\rmd t}} 
&=& \overline{\vb\vb:\vdel\vu}
= \vb_0\vb_0:\vdel\vu_0 
+ \overline{\vb_0\cdot\lt(\vdel\delta\vu_\perp\rt)\cdot\frac{\delta\vB_\perp}{B_0}}\\
&=& \frac{1}{B_0}\frac{\rmd B_0}{\rmd t} + 
\frac{1}{2}\frac{\dd}{\dd t}\overline{\frac{|\delta\vB_\perp|^2}{B_0^2}} 
\approx -\frac{2\nuii}{\beta}, 
\eea 
so the (fast) growth of the firehose fluctuations largely cancels the (slow) 
decrease of the large-scale field and keeps the mean rate of change marginal. 
This is achieved by perturbations both to the field and to the flow. However, 
while their combined effect on $\gamma$ is dramatic, the effect of the perturbations 
just on the rate-of-strain tensor averages out: 
$\overline{\vdel\vu} = \vdel\vu_0 + \overline{\vdel\delta\vu_\perp} = \vdel\vu_0$. 
The rate of strain may be large, but the mean magnetic field does not feel it. 

PIC simulations by \citet{Kunz14} 
of shearing a mean magnetic field in a high-$\beta$ plasma 
provide a numerical example of how the mean rate of change of the magnetic-field 
strength might be reduced (at least transiently) by both firehose and 
mirror fluctuations. Another such example, for a collisional magnetized 
(Braginskii) plasma at the firehose threshold, is reported by \citet{Melville14}. 

\subsection{One-Scale Flow} 
\label{sec:onescale}

Let us first consider a rather artificial situation in which 
the rate of strain $\sigma$ is set without regard to microphysics 
by the model equation \exref{eq:sigma0}. This amounts to assuming 
a fixed viscosity determining a definite cutoff scale for the 
turbulence and, therefore, a definite decorrelation rate, $\sigma_0$, 
for the rate of strain. In reality, the pressure anisotropy
will have a dramatic effect on the viscosity of the ICM 
--- we will take up this further complication 
in \secref{sec:multiscale}.  

\subsubsection{Qualitative Discussion}
\label{sec:M1_qual}

In any given realization, how large the field has managed to grow determines 
how strongly it can be further stretched. This means that in the realizations where the field 
is particularly strong, its growth (or decay) can also be faster. Conversely, in realizations 
where the field is weak, it can neither grow nor decay very fast because higher values 
of $\beta$ constrain the rate of change of the field more stringently. This suggests 
a positive feedback mechanism: consider for a moment a realization 
where $\sigma$ has managed to stay at the mirror threshold at all times: 
$\sigma = B^2\nuii/8\pi p$. Then $\dd_t B = (\nuii/8\pi p)B^3$ and the field growth 
is explosive: 
\beq
B(t) = \frac{B_0}{\sqrt{1-t/t_c}}, \qquad
t_c = \frac{\beta_0}{2\nuii}.
\eeq
Thus, arbitrary field strength can be achieved in finite time.
The condition \exref{eq:sat} for the field to become dynamically 
important is reached at 
\beq
\tdyn = \frac{\beta_0}{2\nuii} - \frac{1}{2\sigma_0}.
\label{eq:t1}
\eeq 
The first term here will typically be much larger than the second and so, 
in comparison with \eqref{eq:t0}, this is a rather sluggish 
field-growth mechanism --- not a surprising outcome as the field 
amplification rate is capped by the mirror threshold.

A further setback for the field growth in this scenario arises from the fact 
that not all realizations of the random stretching rate $\gamma$ will 
manage to keep close to the mirror threshold for a time of order $\tdyn$ 
necessary to consummate the explosive growth. Every time $\gamma$ strays 
into negative values and towards the firehose threshold, the field gets 
weaker, the maximum value of $|\gamma|$ decreases and so recovery and growth become less likely. 
Furthermore, because of the asymmetry of the stable interval of stretching rates, 
$\lt[-2/\beta,1/\beta\rt]\nuii$, 
the decrements in the magnetic energy produced by the negative values of the fluctuating 
rate of strain are on average twice as large as the increments produced 
by its positive values. Thus, there is 
a net tendency for field realizations to decay and any growth will have to come 
from the rare realizations at the mirror end of the distribution 
(see \apref{ap:M1} for a quantitative discussion of this point; 
this tendency is in fact largely an artefact of the zero-dimensional dynamo
model and will be cured in \secref{sec:meanrate}). 

The result of all these effects is that only a small
fraction $f$ of the realizations of the field will manage to grow to 
the dynamical level (i.e., the field will be very poorly space-filling). 
To obtain a crude estimate of $f$, let us estimate the 
probability for a realization to stay at positive values of $\sigma$ (and therefore 
with $\gamma$ at or just below the mirror threshold) throughout the evolution from 
$B_0$ to the dynamical strength \exref{eq:sat}. 
Since the decorrelation rate of $\sigma$ is $\sigma_0$, the rate of strain has 
an opportunity to change sign roughly $\tdyn\sigma_0$ times during the lifetime 
of such a successful realization and the probability for it to stay positive 
each time is $1/2$. Thefore, the fraction of such realizations~is
\beq
f\sim \lt(\frac{1}{2}\rt)^{\tdyn\sigma_0} \sim 
\exp\lt(-\frac{\sigma_0\beta_0}{\nuii}\rt) 
\label{eq:f1}
\eeq
(this is derived in a more quantitative fashion in \apref{ap:M1_tau}). 

\subsubsection{Numerical Results}
\label{sec:M1_num}

We non-dimensionalize our 
equations by letting $M=B^2\nuii/8\pi p\sigma_0 = \nuii/\sigma_0\beta$, $\tt = 2\sigma_0 t$ 
and $\tsigma = \sigma/\sigma_0$. 
Then the equations are 
\bea
\label{eq:M1}
\dd_{\tt} M &=& \lt\{
\begin{array}{cl}
M^2, & \tsigma > M,\\
\tsigma M, & \tsigma \in \lt[-2M, M\rt],\\
-2M^2, & \tsigma < -2M,
\end{array}
\rt.\\
\dd_{\tt} \tsigma &=& -\tsigma + \sqrt{2}\,\chi(\tt). 
\label{eq:sigma1}
\eea
The dynamically-strong-field condition \exref{eq:sat} converts into $M=1$.  
This system depends on no parameters except the initial normalized 
magnetic energy $M_0=\nuii/\sigma_0\beta_0$. 

\Eqsand{eq:M1}{eq:sigma1} are quite straightforward to solve numerically. 
To obtain good statistics, a large of realizations (typically $N=10^7$ to $10^9$) 
were used and a parameter scan in the initial magnetic energy $M_0$ was carried out;
for each value of $M_0$, all realizations had $M(\tt=0)=M_0$, i.e., the initial 
distribution of the magnetic energy was $\delta(M-M_0)$. 

In \figref{fig:M1_tau} (left panel), we show the time $\tt_f$ it takes for a given 
fraction $f$ of the realizations to achieve the dynamical 
level \exref{eq:sat}, $M=1$. At $M_0$ that is not too small, 
all $\tau_f$ follow \eqref{eq:t1}, 
which in dimensionless terms is simply $\ttdyn = 1/M_0 - 1$. 
However, for each $M_0$, there is a maximum fraction of realizations 
$\fmax(M_0)$ that will ever reach $M=1$, while the rest will decay, 
and so $\tt_f\to\infty$ and $f\to\fmax$. The inset in the left panel of \figref{fig:M1_tau} 
shows that $\fmax$ follows the estimate \exref{eq:f1} quite dutifully. 

\begin{figure}
\includegraphics[width=80mm]{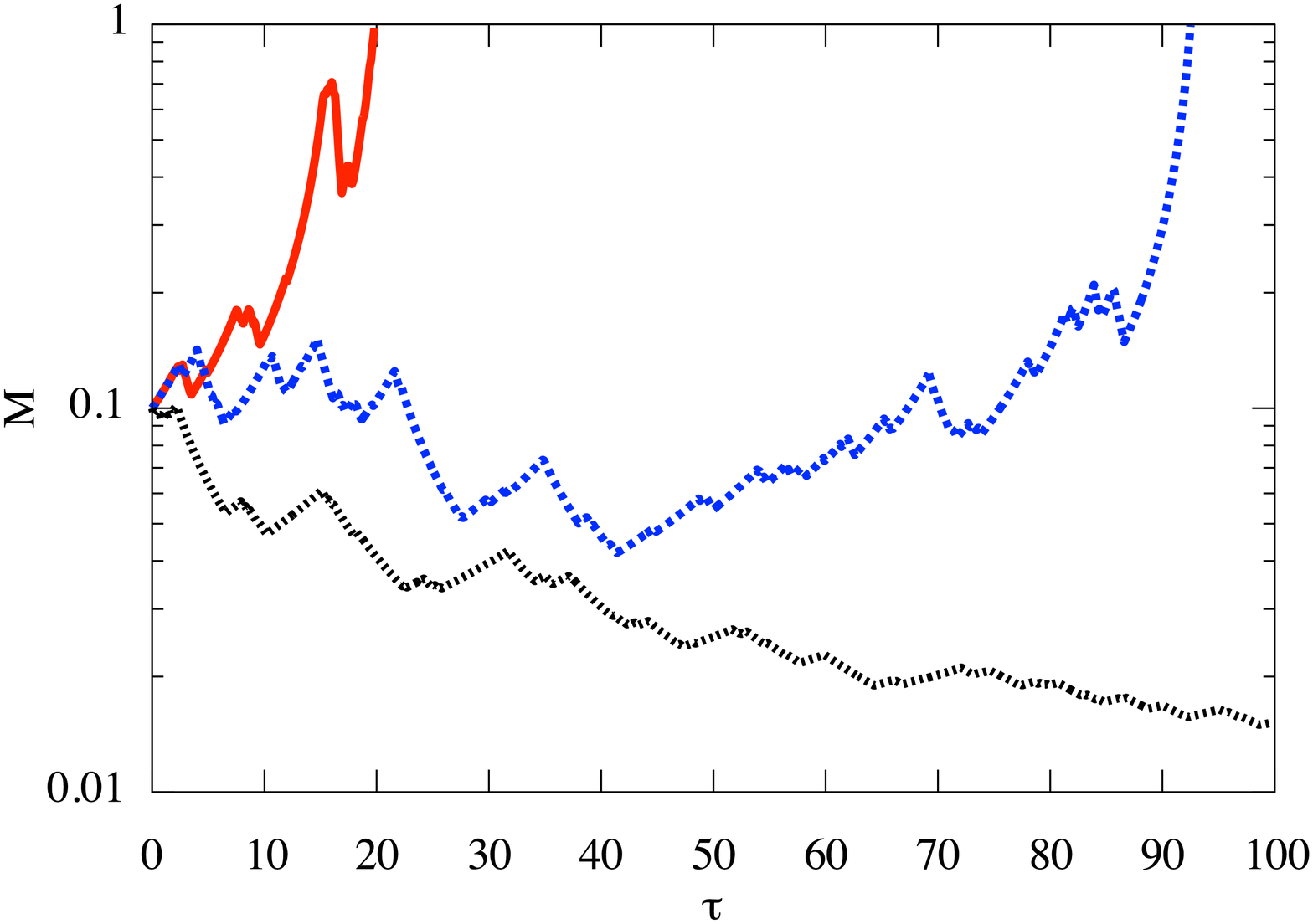}
\caption{Examples of time evolution of growing and decaying realizations in Model~I, 
\eqsand{eq:M1}{eq:sigma1} with $M_0=0.1$. Note the explosive episodes that take 
the field to dynamical level ($M=1$).}
\label{fig:M1_samples} 
\end{figure}

Examples of realizations that grow or decay are shown in \figref{fig:M1_samples}. 
Note that the typical behaviour of a successful dynamo realization is 
explosive growth (on the time scale $\sim\ttdyn$), preceded perhaps 
by a period of hesitation, in line with the qualitative discussion in 
\secref{sec:M1_qual}. \Figref{fig:M1_tau} (right panel) quantifies these 
periods of hesitation in terms of the PDF $P_1(\tt|M_0)$ of the time $\tt$ it takes 
a realization to get from $M=M_0$ to $M=1$. 
The PDF has a peak at $\tt\sim\ttdyn$ and 
an exponential tail for $\tt\gg\ttdyn$
(see \apref{ap:M1_tau} for the derivation of this result).\\   

The conclusion from the above is that getting to dynamically significant 
fields from small initial seeds is both a very slow and a very rare event in 
the scenario we have investigated. 
If a ``dynamo'' is defined by the requirement of growth of mean 
magnetic energy $\la M\ra$, then this is clearly {\em not} a dynamo 
(in \apref{ap:M1}, it is shown that both the typical realizations and all 
moments of $M$ decay). However, there are several ways in which our 
treatment is in fact overly pessimistic and reasonable amendments to 
the model render it much more germane to magnetic-field growth. 
These are discussed in the next two subsections. 

\subsection{One-Scale Flow with a Mean Stretching Rate}
\label{sec:meanrate}

The main reason for the decay of the field in the above treatment 
is that the decay rate at the firehose threshold ($\tsigma=-2M$) 
is larger than the growth rate at the mirror threshold ($\tsigma=M$) 
while the probability for rate of strain to stray beyond either of these thresholds 
is approximately the same when $M\ll1$. This is in fact an artefact of the 
zero-dimensional dynamo model we have adopted (\secref{sec:zeroD}), 
in which it is hard-wired that $\la\sigma\ra = 0$. Therefore, 
in the absence of instabilities, $\la\gamma\ra = 0$ and typical realizations 
neither grow nor decay on average, $\la\ln B\ra=0$; with the instabilities, this gives 
the field decay at firehose threshold a competitive edge. 
In 3D dynamo, this is certainly not the case: the growing field configures 
itself in such a way with respect to the rate-of-strain tensor that 
$\la\gamma\ra = \la\vb\vb:\vdel\vu\ra > 0$ and $\la\ln B\ra = \la\gamma\ra t$ 
(see, e.g., \citealt{Sch04}; 
the value of $\la\gamma\ra$ is related to the mean Lyapunov exponents 
associated with the rate-of-strain tensor; see \citealt{Zeldovich84,Chertkov99,MHDbook}).
Spatially this means that regions where magnetic field decays occupy smaller 
area on average than those where it grows, i.e., the rate of strain is more 
likely to cross the mirror threshold than the firehose one (this was 
quantitatively confirmed in the recent study by \citealt{Lima14}). 

\begin{figure}
\includegraphics[width=80mm]{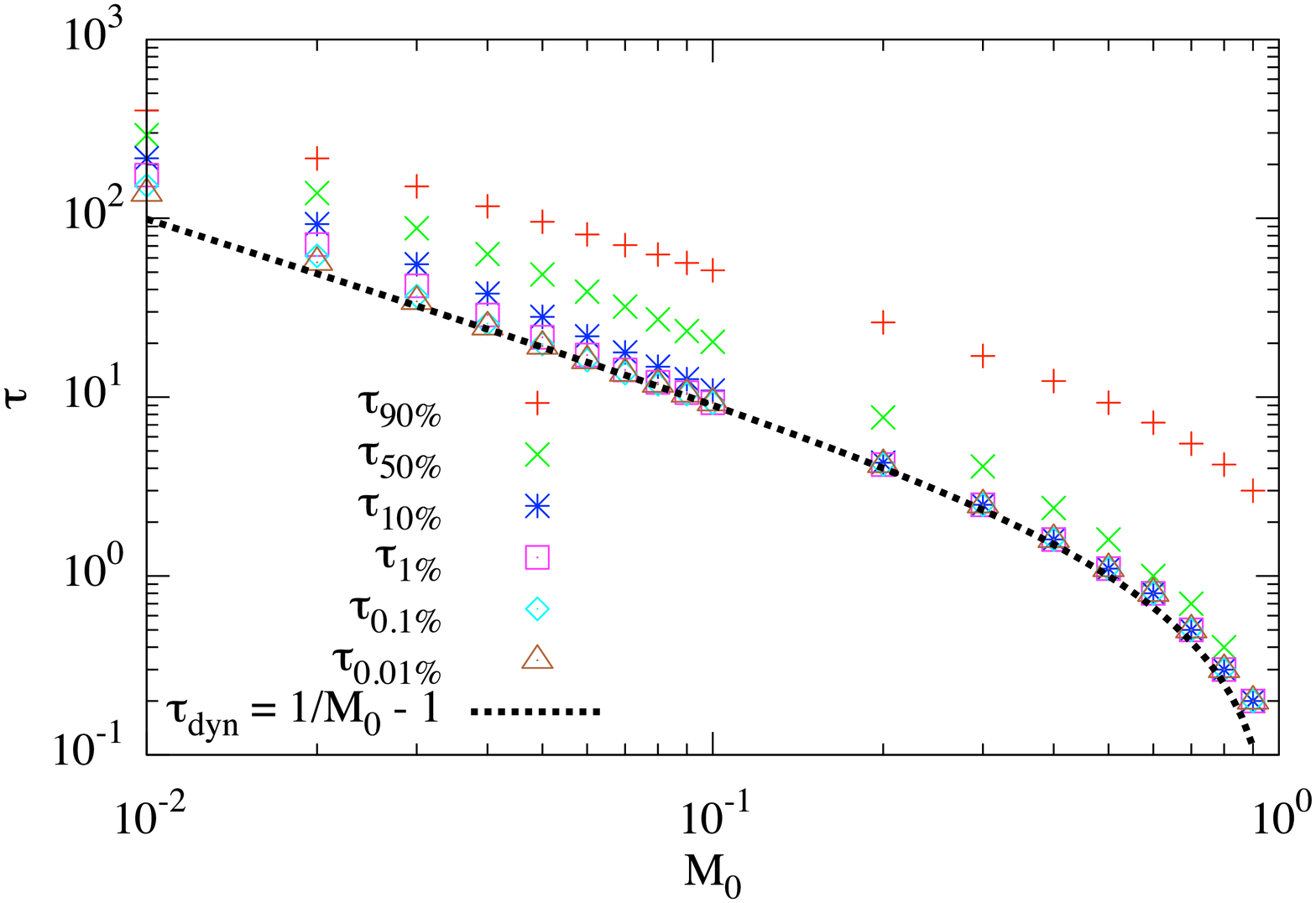}
\caption{Same as \figref{fig:M1_tau} (left panel), but for 
the modified Model~I, \eqsand{eq:M1_gamma}{eq:sigma1} with mean stretching rate $\tgam=0.75$. 
This time all realizations eventually reach $M=1$ on a timescale consistent with \eqref{eq:t1}; 
times $\tt_f$ for $f=90\%$, 50\%, 10\%, 1\%, 0.1\%, and 0.01\% are shown.}  
\label{fig:M1_tau_gamma}
\end{figure}

To incorporate this feature into Model~I, we may stipulate that, 
in the absence of thresholds, $\gamma$ (or, equivalently, $\sigma$) has  
a non-zero prescribed mean $\gamma_0$. 
Then in \eqsand{eq:M1_mirror}{eq:M1_fh}, we replace $\sigma \to \sigma + \gamma_0$. 
\Eqref{eq:M1} is then replaced by     
\beq
\label{eq:M1_gamma}
\dd_{\tt} M = \lt\{
\begin{array}{cl}
M^2, & \tsigma > M - \tgam,\\
\lt(\tsigma + \tgam\rt) M, & \tsigma \in \lt[-2M-\tgam, M-\tgam\rt],\\
-2M^2, & \tsigma < -2M - \tgam,
\end{array}
\rt.
\eeq
where $\tgam = \gamma_0/\sigma_0 > 0$. 

For $M\ll\tgam$, the system is at the mirror threshold with probability 
\beq
p=\frac{1}{\sqrt{2\pi}}\int_{-\tgam}^\infty\rmd\tsigma\,e^{-\tsigma^2/2} >\frac{1}{2}
\label{eq:p_gamma}
\eeq
and at the firehose threshold with probability $1-p<1/2$. Then 
\beq
\lt\la\frac{\dd_\tt M}{M^2}\rt\ra = 3p-2. 
\eeq
The tendency will be for $M$ to decay if $p<2/3$ and to grow if $p>2/3$.  
Using \eqref{eq:p_gamma}, we find that the threshold value corresponds to 
$\tgam\approx 0.4$, but one should not regard this as a quantitative 
prediction for real 3D turbulence because the distribution of the 
rate of strain is in reality very far from being Gaussian. The salient 
point is that the spatial structure of the field and the consequent 
relative space-filling properties of the regions of growth and decay will 
matter. They cannot unfortunately be captured adequately in the zero-dimensional 
modelling framework adopted in this paper and require direct numerical simulations 
of the dynamo saturation in 3D. 

For completeness, an analytical treatment of the Model~I with one-scale 
flow and non-zero mean stretching rate is given in \apref{ap:meanrate}.  
When $p>2/3$, the mean magnetic energy is
\beq
\la M\ra = \frac{1}{1/M_0 - (3p-2)\tt},
\eeq
which explodes at $\tt=1/(3p-2)M_0\sim\ttdyn$ (as do all higher moments of $M$); 
all realizations eventually reach the dynamical level $M=1$, $\fmax=1$. 
The dynamo is explosive and 100\% ``efficient,'' although the growth time 
is still given by \eqref{eq:t1} --- long if the initial field is small. 
\Figref{fig:M1_tau_gamma} confirms this result 
by showing the time it takes for a given fraction of 
realizations to reach $M=1$ in a numerical solution of \eqsand{eq:M1_gamma}{eq:sigma1} 
with $\tgam=0.75$ (so $p\approx 0.77$).

\subsection{Plasma Dynamo and the ICM Viscosity} 
\label{sec:multiscale}

The one-scale-flow dynamo model ignores an essential effect. 
The suppression of the mean field-stretching rate by the microinstabilities 
affects not just the rate of growth of the field but also the effective 
viscosity of the plasma and, therefore, changes the spatial scale of the 
maximum turbulent rate of strain, which is, in Kolmogorov turbulence, the viscous 
cutoff scale. This in turn sets the magnitude of the rate of strain and so 
also its decorrelation rate ($\sigma_0$, thus far assumed fixed). 

To be more quantitative, consider the momentum equation in a magnetized plasma, 
valid at time and spatial scales longer than the Larmor period and radius, 
respectively: 
\beq
\rho\frac{\rmd\vu}{\rmd t} = -\vdel\lt(\pperp + \frac{B^2}{8\pi}\rt)
+ \vdel\cdot\lt[\vb\vb\, p\lt(\Delta + \frac{2}{\beta}\rt)\rt].
\label{eq:u}
\eeq
The mean pressure anisotropy is 
$p\Delta = \mu\,\overline{\vb\vb:\vdel\vu}$, 
where $\mu= p/\nuii$ is the (parallel) viscosity of the plasma
(see \eqsand{eq:Delta}{eq:B}). 
In the absence of the microinstabilities, it is this term 
that provides the viscous damping of the component of the rate 
of strain that can change the magnetic field strength (the stretching rate), 
while the $2/\beta$ term is the tension force (Maxwell stress), 
responsible for back reaction of the field on the flow --- this 
back reaction was assumed small for weak fields being amplified by a dynamo. 
Since, under our modelling assumptions, $\overline{\vb\vb:\vdel\vu}$ is 
suppressed by the instabilities and so $|\Delta|$ is never larger 
than $2/\beta$, there is no mechanism left to enforce the viscous cutoff
on any part of the rate-of-strain tensor. 
Thus, as the system crosses the stability thresholds \exref{eq:interval}, 
the turbulent cascade is free to extend to very small scales, probably 
into the microscale range where the finite-Larmor-radius (FLR) 
effects (omitted in \eqref{eq:u}) determine the shape of the velocity spectrum 
and where also the firehose and mirror instabilities operate, so  
scale separation is lost between mascroscopic motions and the microphysics. 

As this happens, the maximum rate of strain $\sigma$ becomes very large 
even as the velocity of the motions becomes very small ($\sigma\propto l^{-2/3}$ 
but $u\propto l^{1/3}$ in Kolmogorov turbulence; $l$ is scale). 
Therefore, already very weak magnetic fields 
can exert dynamical back reaction on these motions and, if the field is sufficient 
to do that, the smallest scale at which the motions are capable of amplifying the field 
will be the scale where the kinetic-energy density of the motions is 
comparable to the total magnetic-energy density, $\rho u_l^2/2 \sim B^2/8\pi$. 
Since $u_l\sim (\eps l/\rho)^{1/3}$, this gives 
\beq
l \sim \frac{p\,\vthi}{\eps}\,\beta^{-3/2},
\label{eq:lB}
\eeq
where $\vthi = (p/\rho)^{1/2}$ is the ion thermal speed. 
The corresponding rate of strain is
\beq
\sigma \sim \frac{u_l}{l} \sim \frac{\eps}{p}\,\beta \sim \frac{\sigma_0}{M}. 
\label{eq:sigma_lB}
\eeq
As the magnetic field grows, the scale \exref{eq:lB} gets larger and the rate of strain \exref{eq:sigma_lB} 
smaller. Note, however, that the rate at which the field is amplified is at all times 
limited by the instabilities to be $\lesssim\nuii/\beta\ll\sigma$, as given by \eqref{eq:M1_mirror},
so \eqref{eq:sigma_lB} just gives the decorrelation rate of the stretching, not its 
actual magnitude, i.e., the inverse of the typical time that the system might 
spend at the mirror threshold before flipping the sign of $\sigma$ and 
ending up at the firehose threshold. 

The argument involving the tension force limiting the motions 
does not apply at the firehose threshold: indeed, there $\Delta + 2/\beta = 0$ 
and so the viscous damping and the nonlinear back reaction in 
\eqref{eq:u} exactly cancel each other: field lines lose tension 
and the motions no longer feel them at all. This means that 
in the regions and instances where the field weakens, the cutoff scale for the motions 
is microphysical (FLR-determined), giving the rate of strain --- and, therefore, the decorrelation 
rate --- much larger than in the growing-field regions.

Thus, as the field grows, the typical time the system spends at 
the mirror threshold becomes much larger than the time it spends at 
the firehose threshold. This resembles the dynamo discussed 
in \secref{sec:meanrate}: $\dd_\tt M = M$ with probability $p\to1$ 
and $\dd_\tt M = -2M$ with probability $1-p\to0$. The result will, 
therefore, be a robust explosive dynamo with the typical growth time again 
given by \eqref{eq:t1}.   

One can construct model dynamical equations for $\sigma$ 
(to replace \eqref{eq:sigma1}) that would include all of the above effects 
and then solve them numerically together with \eqref{eq:M1}.\footnote{The treatment 
of the enhanced-collisionality case in \secref{sec:M2} is an example of how 
nonlinear equations for $\sigma$ incorporating dependence on $M$ might be constructed.} 
However, the level of uncertainty about 
the way in which the rates of strain at the mirror and, especially, 
firehose thresholds are determined is such that a zero-dimensional 
modelling exercise is unlikely to teach us much more than the above qualitative 
discussion has done --- and any further conclusions will be sensitive 
to a large number of modelling choices. A more promising course of action here 
would be direct 3D numerical simulations using Braginskii MHD 
equations with a suitable implementation of the microphysical closure 
corresponding to Model~I --- such a study, although  
highly desirable, is outside the scope of this paper.  

\subsection{Constraints on the Seed Field}
\label{sec:M1_constraints}

In order for the above considerations to be applicable, time- and 
spatial-scale separation between macro- and microphysics must be present. 
This imposes a number of constraints all of which can be expressed as lower 
bounds on the magnitude of the magnetic field --- and, therefore, on the initial 
seed field from which any of the dynamo models considered so far is allowed to start. 

These constraints can be expressed in terms of $\beta$ (or $M$) and 
the three relevant time scales in the problem, for which we will 
adopt reference core ICM values \citep[cf.][]{Sch06,Ensslin06,Rosin11,Kunz11}:  
\bea
\label{eq:ICM_sigma}
\sigma_0 &\sim& 10^{-14}~{\rm s}^{-1},\\
\nuii &\sim& 10^{-12}~{\rm s}^{-1},\\
\Omega_1 &\sim& 1~{\rm s}^{-1},
\label{eq:ICM_Omega}
\eea
where $\Omega_1 = e\sqrt{8\pi p}/mc$ is the Larmor frequency corresponding to $\beta=1$.  
Let us itemize the constraints.\\ 

\noindent (i) Plasma must be magnetized: $\nuii\ll\Omega_i$, provided
\beq
\frac{1}{\beta} \gg\lt(\frac{\nuii}{\Omega_1}\rt)^2 \sim 10^{-24},
\ {\rm or}\ 
M \gg \frac{\nuii^3}{\sigma_0\Omega_1^2}\sim 10^{-22}, 
\label{eq:magnetized}
\eeq
a constraint overriden by the more stringent ones to come.\\

\noindent (ii) The typical growth rates of the mirror and firehose instabilities 
must be larger than the turbulent rate of strain in order for an ``instantaneous'' 
suppression of the latter to be a sensible assumption. 
The peak growth rates of the parallel firehose \citep[e.g.,][]{Davidson68,Sch10} 
and mirror \citep{Hellinger07} are
\bea
\label{eq:gfir}
\gfir &\sim& \Omega_i\lt|\Delta + \frac{2}{\beta}\rt| \sim \frac{\Omega_1}{\beta^{3/2}},\\
\gmir &\sim& \Omega_i\beta\lt(\Delta - \frac{1}{\beta}\rt)^2 \sim \frac{\Omega_1}{\beta^{3/2}}.
\label{eq:gmir}
\eea
For the purposes 
of these estimates, we take $\Delta\sim1/\beta$ and assume that the distance 
to threshold is also of order $1/\beta$ (assuming $\beta\gg1$).
This is reasonable \citep[e.g.,][]{Rosin11}, but not necessarily obvious
(especially for the mirror; see \citealt{Hellinger07} --- but we do not have 
a better {\em a priori} estimate).
If the microinstabilities, in their nonlinear state, manage to keep 
the pressure anisotropy even more tightly pinned to the threshold, 
their effective growth rates become smaller, scales larger and 
the resulting lower bounds on the seed field even more stringent. 
Note that the oblique firehose grows faster than the parallel 
one \citep{Yoon93,Hellinger00}, so if $\gfir$ and $\gmir$ given by 
\eqsand{eq:gfir}{eq:gmir} are large enough, so will be the growth rate 
of the oblique firehose. 

Thus, the reference instability growth scale we will use is $\gmir\sim\Omega_1/\beta^{3/2}$. 
Then $\sigma_0\ll\gmir$ if
\beq
\frac{1}{\beta} \gg \lt(\frac{\sigma_0}{\Omega_1}\rt)^{2/3}\sim 10^{-10},
\ {\rm or}\ 
M \gg \frac{\nuii}{\sigma_0^{1/3}\Omega_1^{2/3}}\sim 10^{-8}. 
\eeq
With the heuristic model of the ICM viscosity proposed in \secref{sec:multiscale}, 
the effective rate of strain at the mirror threshold is given by \eqref{eq:sigma_lB} 
and so $\sigma\ll\gmir$ if
\beq
\frac{1}{\beta} \gg \lt(\frac{\eps}{p\Omega_1}\rt)^{2/5}\sim 10^{-7},
\ {\rm or}\ 
M \gg \frac{\nuii^{3/5}}{\sigma_0^{1/5}\Omega_1^{2/5}}\sim 10^{-5} 
\label{eq:gmir_bound}
\eeq
(we have used $\eps/p = \sigma_0^2/\nuii$). 
Note that $\gfir$ might be a sensible estimate for the maximum rate 
of strain accessible at the firehose threshold, in which case the 
condition for the system to be at the mirror threshold with larger 
probability ($p \sim 1 - \sigma/\gfir$) than at the firehose threshold 
is $\sigma \ll \gfir$, which is automatically ensured by \eqref{eq:gmir_bound}.\\ 

\noindent (iii) The typical scales of the mirror and firehose instabilities 
must be shorter than the scale of the motions that stretch the magnetic field. 
The (parallel) scales at which the peak growth rates \exref{eq:gfir} and \exref{eq:gmir} 
are achieved are 
\bea
\lfir &\sim& \frac{\rho_i}{\sqrt{\lt|\Delta + 2/\beta\rt|}} \sim \frac{\vthi}{\Omega_1}\beta,\\
\lmir &\sim& \frac{\rho_i}{\beta(\Delta - 1/\beta)} \sim \rho_i \sim \frac{\vthi}{\Omega_1}\beta^{1/2}, 
\eea
where $\rho_i=\vthi/\Omega_i$ is the Larmor radius. 
For the oblique firehose, $\lfir\sim\rho_i$. 
The constraints that are obtained by requiring $l\gg\lmir,\lfir$ 
are less stringent than \eqref{eq:gmir_bound}.\\ 

\noindent (iv) The use of \eqref{eq:M1_mirror} for the effective rate 
of amplification of the magnetic field is 
predicated on the calculation of the pressure anisotropy from the balance 
of the rate of change of the field and the collisional isotropization 
(see \eqref{eq:Delta}), which requires the collision time scale to be shorter than 
any fluid time scales. On the other hand, the collision rate must be 
smaller than the growth rates of the instabilities. 
Requiring $\nuii\ll\gmir$ gives 
\beq
\frac{1}{\beta} \gg \lt(\frac{\nuii}{\Omega_1}\rt)^{2/3}\sim 10^{-8},
\ {\rm or}\ 
M \gg \frac{\nuii^{5/3}}{\sigma_0\Omega_1^{2/3}}\sim 10^{-6}. 
\eeq

In \secref{sec:multiscale}, the rate 
of strain becomes large for small $M$ (see \eqref{eq:sigma_lB}) and so 
the requirement $\sigma\ll\nuii$ imposes what turns out to be the most stringent 
of the lower bounds on the magnetic energy: 
\beq
\frac{1}{\beta} \gg \frac{\eps}{p\nuii}\sim 10^{-4},
\ {\rm or}\ 
M \gg \frac{\sigma_0}{\nuii} \sim 10^{-2}. 
\label{eq:coll_bound}
\eeq
If this is violated, further modelling choices have to be made regarding 
the handling of the collisionless mechanism for setting the relationship between 
the mean pressure anisotropy and the rate of change of the magnetic field,\footnote{We stress
that it is the relationship between $\Delta$ and $\gamma$ that is unclear; 
the pressure anisotropy itself is likely always to be at the mirror or 
firehose theresholds \citep{Kasper02,Hellinger06,Bale09,Laveder11}.} 
currently a poorly understood issue, which we leave outside 
the scope of this paper and which is likely to require some form of collisionless 
Landau-fluid closure \citep[e.g.,][]{Snyder97,Passot07,Passot12}. 

\subsection{Summary for Model~I} 
\label{sec:M1_sum}

In this model, the pressure anisotropy is kept from crossing the firehose 
and mirror thresholds by the effective suppression of the mean rate of 
change of the magnetic field. This means that the field growth is 
generally less efficient than it would have been without the 
microinstabilities. While the growth is, mathematically speaking, 
explosive, the time for it to happen is $\tdyn\sim\beta_0/\nuii$ 
(\eqref{eq:t1}), which can be very long if the initial seed is small 
and collisions are rare. 

If we accept this scenario, the implication is that it is virtually 
impossible to generate fields of observed strength ($B\sim10^{-6}$~G, 
or $\beta\sim 10^2$, or magnetic energy $M\sim 1$ in our dimensionless terms) 
from purely primordial seeds ($B\sim10^{-21}-10^{-9}$~G; see review by \citealt{Durrer13}). 
If, on the other hand, there is (or was) a sustained source of field only 
one or two orders of magnitude below the dynamical strength, then the dynamical 
field level can perhaps be effectively maintained by turbulence, possibly in very 
intermittent patches. To study this, one would need to combine a dynamo saturation 
model \citep[e.g.,][]{Boldyrev01,Sch02} with the model of field evolution 
in the presence of pressure anisotropies. Given the large amount of theoretical 
uncertainty in the understanding of both, we leave this outside the scope 
of this paper --- and note that significant progress could be made via 
direct numerical simulations incorporating a microphysical closure 
represented by Model~I. 

An essential caveat to the above conclusion is that all versions of Model~I 
that we have discussed are valid only for relatively large magnetic fields, owing 
to a number of lower bounds on the magnetic energy imposed by the 
requirement of scale separation between fluid, collisional and 
microinstability time and spatial scales (\secref{sec:M1_constraints}). 
The most pessimistic of these bounds (\eqref{eq:coll_bound}) constrains 
$B$ to values only an order of magnitude below the target dynamical 
strength and thus means that these ideas are primarily useful for the study 
of how the field is structured and maintained in the currently observed 
turbulent ICM rather than how it grew to its present level from a tiny seed. 
The question of magnetogenesis starting from a tiny seed may have to wait 
for a better understanding of field growth in a collisionless 
(as opposed to weakly collisional) plasma (see further discussion in 
\secref{sec:maggen}). 

\section{Model II: Enhanced collisionality}
\label{sec:M2}

\subsection{Microphysical Closure}

Now we examine the possibility that it is not the stretching rate of the field 
but the effective collisionality of the plasma that keeps the pressure anisotropy 
from crossing the instability thresholds. 
The stretching rate is always $\gamma=\sigma$ and the rate of strain $\sigma$ continues 
to obey \eqref{eq:sigma0} as long as $\sigma\in\lt[-2,1\rt]\nuii/\beta$, but 
whenever it falls outside this interval, we postulate an instantaneous 
adjustment of the collision rate: 
\beq
\nueff = \xi|\sigma|\beta,
\label{eq:nueff}
\eeq
where $\xi=1/2$ and $1$ at the firehose and mirror thresholds, respectively. 
Larger collision frequency means smaller effective viscosity, 
$\mueff = p/\nueff = \mu\nuii/\nueff$, where $\mu$ and $\nuii$ continue to 
denote the ``bare'' viscosity and collisionality associated with Coulomb collisions. 
Therefore, locally, the Kolmogorov cascade will extend to smaller scales and 
larger rates of strain $\sim(\eps/\mueff)^{1/2}$ \citep{Sch06}. 
We can model this by replacing in \eqref{eq:sigma0}
\beq
\sigma_0 \to \sigma_0\lt(\frac{\nueff}{\nuii}\rt)^{1/2} 
= \sigma_0\lt(\frac{\xi|\sigma|\beta}{\nuii}\rt)^{1/2}, 
\label{eq:sigmaeff}
\eeq
where $\sigma_0$ will continue to denote the ``bare'' rms rate of strain. 
Adopting again the non-dimensionalization introduced in \secref{sec:M1_num}, 
we replace \eqsand{eq:M1}{eq:sigma1} 
with\footnote{We use the Stratonovich stochastic calculus discretization rule 
for the term containing white noise in \eqref{eq:sigma2}, i.e., 
$\dd_{\tt}\tsigma = [\sigma(\tt+\rmd \tt) - \sigma(\tt)]/\rmd \tt$
and $\tsigma$ in the right-hand side is $[\tsigma(\tt+\rmd\tt) + \tsigma(\tt)]/2$; 
$\chi(\tt) \rmd \tt$ is the Wiener measure. The numerical results obtained 
using the It\^o calculus instead are not significantly different.} 
\bea
\label{eq:M2}
\dd_{\tt} M &=& \tsigma M,\\
\label{eq:sigma2_stable}
\dd_{\tt} \tsigma &=& -\tsigma + \sqrt{2}\,\chi(\tt),
\qquad\qquad\qquad\qquad\quad\ 
\xi|\tsigma|\le M.\\ 
\dd_{\tt} \tsigma &=& -\lt(\frac{\xi|\tsigma|}{M}\rt)^{1/2}\!\!\!\tsigma 
+ \sqrt{2}\lt(\frac{\xi|\tsigma|}{M}\rt)^{3/4}\!\!\!\chi(\tt),
\ \, 
\xi|\tsigma|>M. 
\label{eq:sigma2}
\eea

Note that our approach here differs from the earlier work by \citet{Sch06} (where 
the possibility of an explosive dynamo was first explored) in that there a specific 
microphysically inspired formula for the effective collisionality was 
postulated, while here we take a more agnostic attitude and simply assume that the 
anisotropy will always be effectively pinned at the marginal level. 
In this sense, the model proposed here is a more adequate reflection of what ought 
to happen in numerical simulations that adopt the prescription of sharply increased 
local collisionality to prevent firehose and mirror instabilities from 
developing \citep{Sharma06,Meng12,Kunz12,Lima14}. However, in none of these simulations 
has it so far been possible to accommodate numerically the dramatic local refinement of 
the viscous dissipation scale --- the key effect here! --- and thus we remain in the 
realm of largely unexplored physics (see discussion in \secref{sec:modelling}). 

\subsection{Qualitative Discussion}
\label{sec:M2qual}

Ignoring for the purposes of a quick estimate the difference between the rate of strain 
$\sigma$ and its rms value, we find from \eqref{eq:sigmaeff}
\beq
\sigma \sim \frac{\sigma_0^2}{\nuii}\,\beta \sim\frac{\eps}{p}\,\beta\sim \frac{\sigma_0}{M}
\quad\Rightarrow\quad
\mueff \sim \mu\frac{\nuii}{\nueff} \sim \mu M^2, 
\label{eq:sigmaest}
\eeq
so realizations with weaker magnetic field will have smaller viscosity, 
faster stretching, smaller viscous scale (Kolmogorov scale $\propto\mueff^{3/4}$), 
but also smaller velocities at this scale ($\propto\mueff^{1/4}$). This means 
that, with respect to these velocities, the magnetic field will not need to be 
very strong in order to start having a dynamical effect --- and examining the 
condition \exref{eq:sat} with $\mu$ replaced by $\mueff$, we discover that 
it is in fact exactly satisfied by the estimates \exref{eq:sigmaest}. 
Thus, in this scenario, the dynamo becomes nonlinear the moment either of the instability 
thresholds is crossed. 

It might appear that this development requires some amendment 
to our model reflecting the role of the newly ascendant 
Lorentz back reaction in moderating the dynamo. In fact, our model already 
takes care of this effect, on a qualitative level. 
Indeed, let us consider what happens when the magnetic energy density becomes 
comparable to the kinetic energy density of the viscous-scale turbulent ``eddies''. 
These eddies (or, to be precise, the stretching rate 
associted with them, $\vb\vb:\vdel\vu$) become suppressed and 
the field is now stretched by the eddies at the next largest scale, 
grows to have energy density comparable to the eddies at that scale, suppresses 
them, and so on until it has thus worked its way up the inertial range to be 
in energy equipartition with the largest turbulent motions. This scenario 
\citep{Sch02} was recently studied and validated in the MHD 
numerical simulations of \citet{Cho09,Beresnyak12} (ideologically it goes back to the 
classic paper by \citealt{Biermann51}). 
The field growth can be modelled by requiring that magnetic energy, as it grows, 
is always amplified by the turbulent eddies $u_l$ at scale $l$ 
that have the same energy density: 
\beq
\frac{\rmd}{\rmd t} B^2 \sim \frac{u_l}{l}\,B^2 \sim \frac{\rho u_l^3}{l} \sim \eps 
\quad\Rightarrow\quad B^2 \sim \eps t, 
\label{eq:ndynamo}
\eeq
where $\eps$ is the constant Kolmogorov cascade rate. 
This is precisely what would happen in our model except the scale $l$ will 
always be the viscous scale set by the effective viscosity $\mueff$: indeed, 
using \eqref{eq:sigmaest}, $\tsigma\sim 1/M$, in \eqref{eq:M2}, we get 
\beq
\dd_{\tt}M \sim 1,
\label{eq:Msecular}
\eeq
and so the magnetic field reaches $M=1$ (which corresponds 
to dynamical strength with respect to the eddies at the viscous scale 
set by the ``bare'' Coulomb collision rate; see \eqref{eq:sat}) at the time
\beq
\ttdyn \sim 1-M_0,\quad{\rm or}\quad
\tdyn \sim \frac{1}{\sigma_0}\lt(1 - \frac{\nuii}{\sigma_0\beta_0}\rt).
\label{eq:t2}
\eeq 
This is at most one large-scale stretching time, regardless of the size of the initial 
field --- a much faster dynamo than achieved by Model I (\eqref{eq:t1}) or 
by the conventional MHD dynamo (\eqref{eq:t0}).  

\subsection{Numerical Results: Effects of Randomness and Finite 
Decorrelation Rates}

\begin{figure}
\includegraphics[width=80mm]{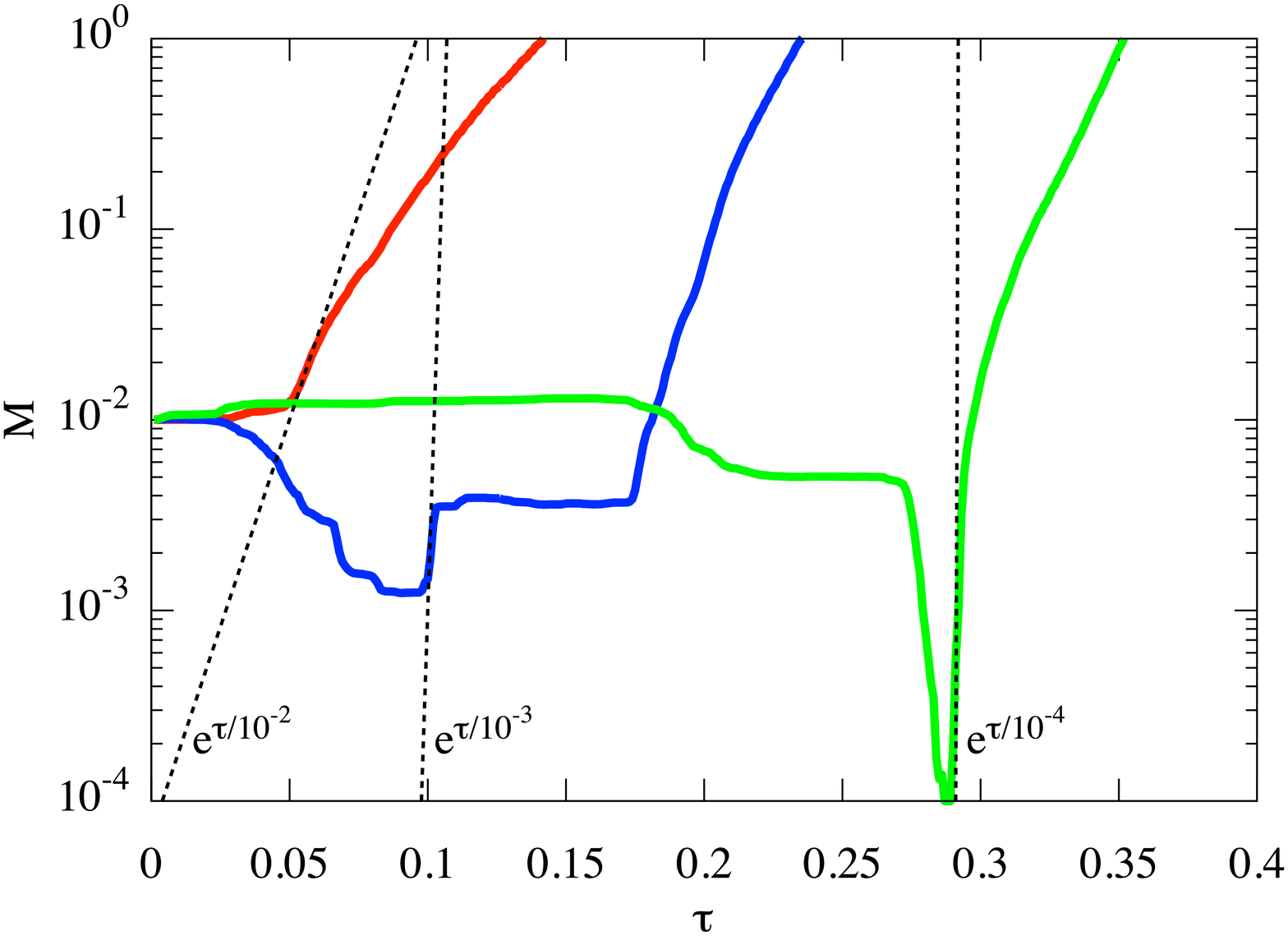}
\caption{Examples of the time evolution of particular realizations in Model~II, 
\eqsdash{eq:M2}{eq:sigma2} with $M_0=0.01$. Episodes of intense growth are manifest here; 
this growth is exponential in time with the rate $1/M$, where $M$ is the magnetic energy at the 
beginning of the episode --- the corresponding slopes are shown as dotted lines. 
See discussion in \secsand{sec:M2growth}{sec:M2time}.}
\label{fig:M2_samples} 
\end{figure}

\begin{figure*}
\includegraphics[width=85mm]{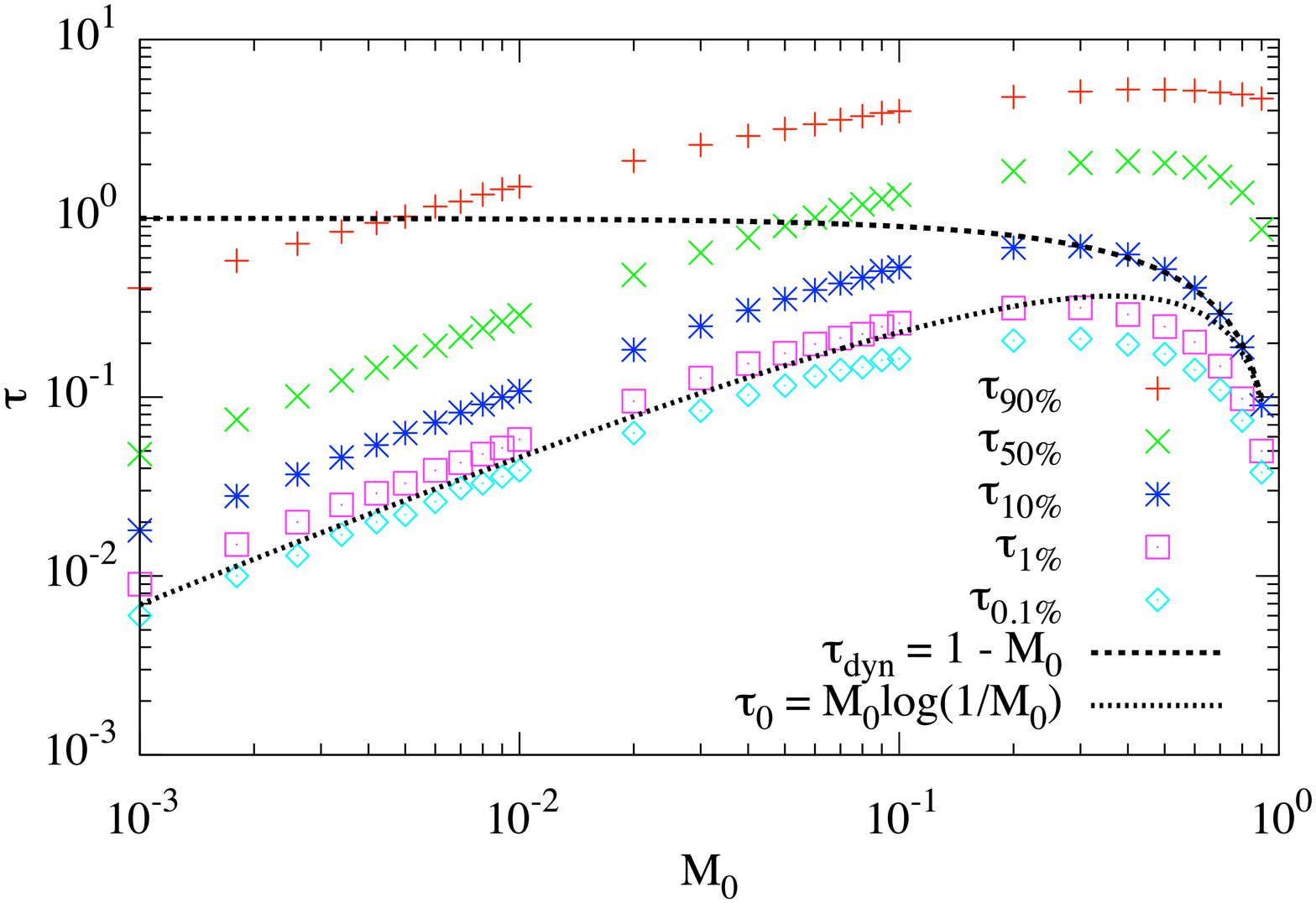}
\includegraphics[width=85mm]{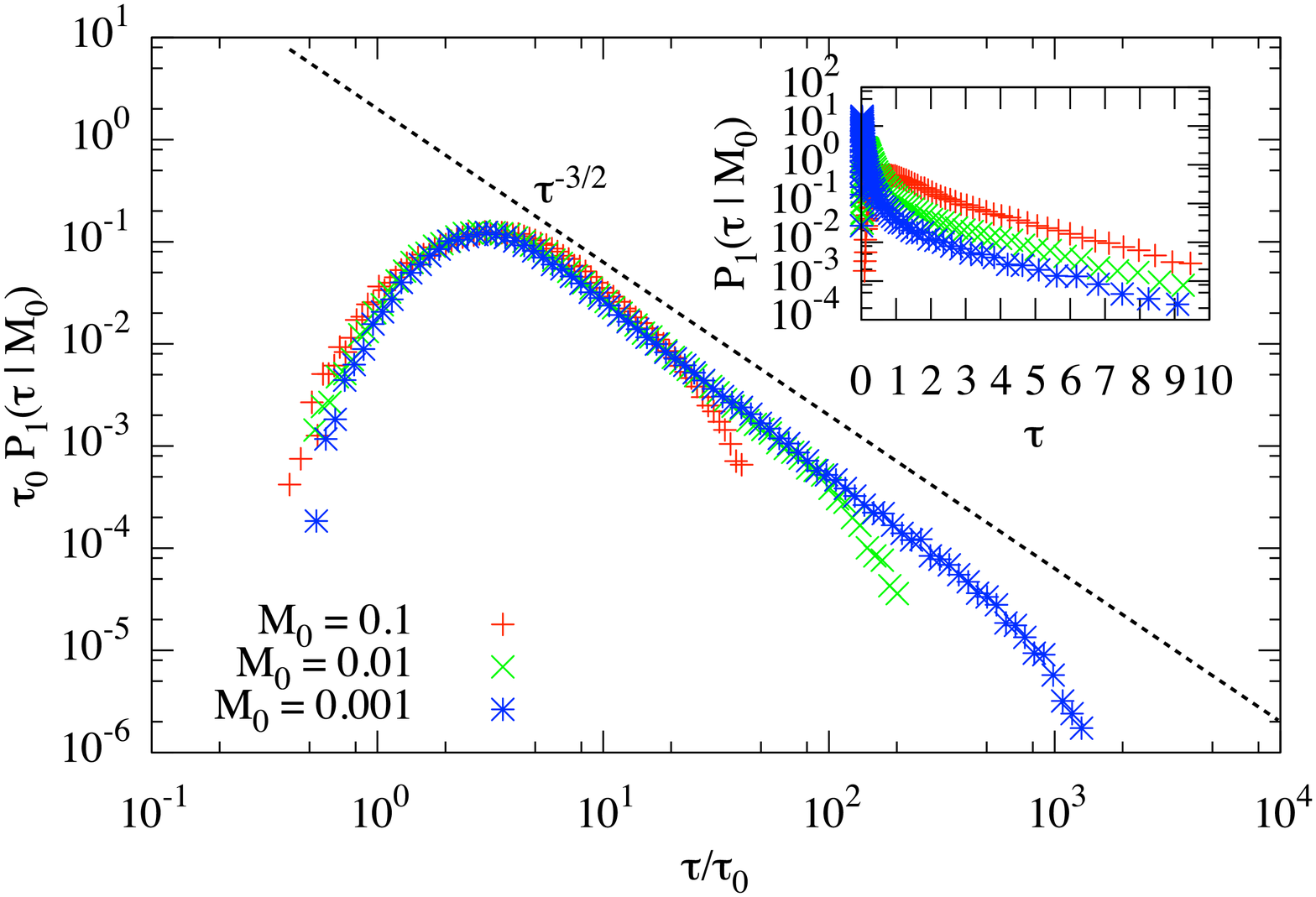}
\caption{{\it Left panel:} Times $\tau_f$ for the fractions $f=90\%$, 50\%, 10\%, 1\%, and 0.1\%, 
of all realizations to achieve $M=1$ in Model~II, 
\eqsdash{eq:M2}{eq:sigma2}, plotted vs.\ initial normalized field energy $M_0=\nuii/\sigma_0\beta_0$. 
The two dotted lines are $\ttdyn = 1 - M_0$ (\eqref{eq:t2}) and $\tt_0=M_0\ln(1/M_0)$ 
(\eqref{eq:t2_fast}). 
{\it Right panel:} PDF of the time $\tt$ it took those realizations to 
grow to $M=1$. Three cases, $M_0=0.1$, $0.01$, $0.001$ are shown here, with $\tt$ 
rescaled by $\tt_0$. The slope corresponding to $\tt^{-3/2}$ is shown for reference. 
{\em Inset:} Same PDFs, but without rescaling by $\tt_0$ and shown on 
a log-linear plot to highlight the exponential cutoff at $\tt\gtrsim1\sim\ttdyn$.} 
\label{fig:M2_tau}
\end{figure*}

\subsubsection{Why the Field Grows}
\label{sec:M2growth}

In view of the experience of a dynamo failure in the simplest 
version of Model~I (\secref{sec:onescale}), we ought to ask why the above scenario 
should preferentially produce growth of the field rather than decay. 
The key difference between Model~II and Model~I is that in Model~II, 
the modification of the field-stretching rate as a result of crossing the
instability thresholds is not instantaneous: it is the effective collisionality $\nueff$ 
that is modified on the short time scales associated with the instabilities, 
which then leads to the change in the instantaneous decorrelation rate of $\tsigma$
(see \eqref{eq:sigma2}). The actual value of $\tsigma$ then takes 
the time $\tt\sim(|\tsigma|/M)^{-1/2}$ to change to a new random value, 
with a modified rms of order $1/M$. 

Consider the evolution of any particular realization of the field. 
If it finds itself at the mirror threshold, the field grows, 
with it increases the effective collisionality, so the instantaneous 
decorrelation rate of $\tsigma$ decreases and the system can spend a 
longer time at this threshold before $\tsigma$ flips sign. 
In contrast, at the firehose threshold, the field drops, 
the decorrelation rate goes up and so the system can revert to 
positive growth rate sooner --- and when it does, it grows at 
a relatively higher rate, $\tsigma\sim1/M$, because it starts 
at a lower value of $M$.\footnote{Note that in truth, the system's 
reluctance to linger at the firehose threshold may be even greater than 
our model allows because the exact cancellation of the viscous stress 
will likely drive the effective decorrelation rate even higher. This 
is reminiscent of the discussion in \secref{sec:multiscale} except
it is the enhanced collisionality that sets $\Delta + 2/\beta=0$ this time, 
so the rate of change of the field is not instantly affected --- and by the 
time it is affected, it may have changed sign. We do not attempt to model 
the effect of the cancellation of the viscous stress 
because the enhanced collisionality produced by the 
firehose fluctuations during an episode of field decay may actually 
set the expectation value of the growth rate after one decorrelation time.} 

\Figref{fig:M2_samples} shows examples of time histories $M(\tt)$ 
obtained by numerical solution of \eqsdash{eq:M2}{eq:sigma2}. 
A particularly striking feature that is manifest here is that the 
field growth can happen in short intense bursts during which the rate 
of increase is exponential and so much faster than suggested by the ``nonlinear-dynamo'' 
secular-growth estimate~\exref{eq:ndynamo}. These fast-growth episodes 
allow the field to reach $M=1$ much more quickly 
than predicted by the estimate \exref{eq:t2}.

\subsubsection{Time to Saturation}
\label{sec:M2time}

Consider a realization starting with $M=M_0\ll1$ and quickly finding itself at the 
mirror threshold. Then the initial growth rate is $\tsigma \sim 1/M_0$ (\eqref{eq:sigmaest}) 
and so $M\sim M_0\exp(\tt/M_0)$. Allowing for a (relatively rare) instance in which this 
value of $\tsigma$ persisted longer than the typical correlation time 
(of order $M_0$ initially, but getting longer as $M$ grows), 
we find that $M=1$ is achieved after 
\beq
\tt_0\sim M_0\ln\lt(\frac{1}{M_0}\rt), 
\quad {\rm or}\quad  
t_0 \sim \frac{p}{\eps\beta_0}\ln\lt(\frac{\sigma_0\beta_0}{\nuii}\rt) \ll \tdyn. 
\label{eq:t2_fast}
\eeq
This means that some number of realizations 
will get there much earlier than suggested by the estimate \exref{eq:t2}. 
These are the fast-growth episodes seen in \figref{fig:M2_samples}. 
%(note that they are even faster if preceded by a period of decay because $M$ is lower). 

\Figref{fig:M2_tau} quantifies their contribution the overall field growth using the numerical 
solution of \eqsdash{eq:M2}{eq:sigma2} (typically for $N=10^5$ realizations). 
The left panel shows that the time 
$\tt_f$ needed for a given fraction $f$ of these realizations 
to achieve $M=1$ actually decreases at small $M_0$, in contrast to the more conservative 
prediction \exref{eq:t2}. 
The PDF of the time for the realizations starting at $M=M_0$ to reach $M=1$, $P_1(\tt|M_0)$ (right panel), 
has a peak around $\tt=\tt_0$, 
followed by a power-law tail with a scaling that appears to be $\sim\tt^{-3/2}$
(cf.\ the MHD case, \eqref{eq:P1_0D}), 
and then an exponential cutoff at $\tt\gtrsim1$ 
(i.e., $t\gtrsim\tdyn$, as per \eqref{eq:t2}; see the inset in the left panel of \figref{fig:M2_tau}). 
Eventually all realizations reach $M=1$ (the dynamo is 100\% efficient), 
with a (relatively small) fraction arriving much earlier than others
(of order 10\%; these are the realizations with $\tt$ up to the left of 
the peak of the PDF). 

\subsubsection{Stochastic Nonlinear Plasma Dynamo}
\label{sec:M2stoch}
 
It is interesting, for completeness, to examine what happens at times 
$\tt_0\ll\tt\ll\ttdyn\sim1$, i.e., during the period when the system settles 
into self-similar evolution, as suggested by the power-law behaviour of $P_1(\tt|M_0)$.
Since at this point we are getting deeper into the study of the precise properties 
of the particular zero-dimensional model we have chosen (\eqsdash{eq:M2}{eq:sigma2}) 
--- properties that may or may not carry over quantitatively to the more realistic situations,   
--- we have exiled the more detailed treatment to \apref{ap:M2}. 
Let us discuss its results in qualitative terms. 

\begin{figure*}
\includegraphics[width=85mm]{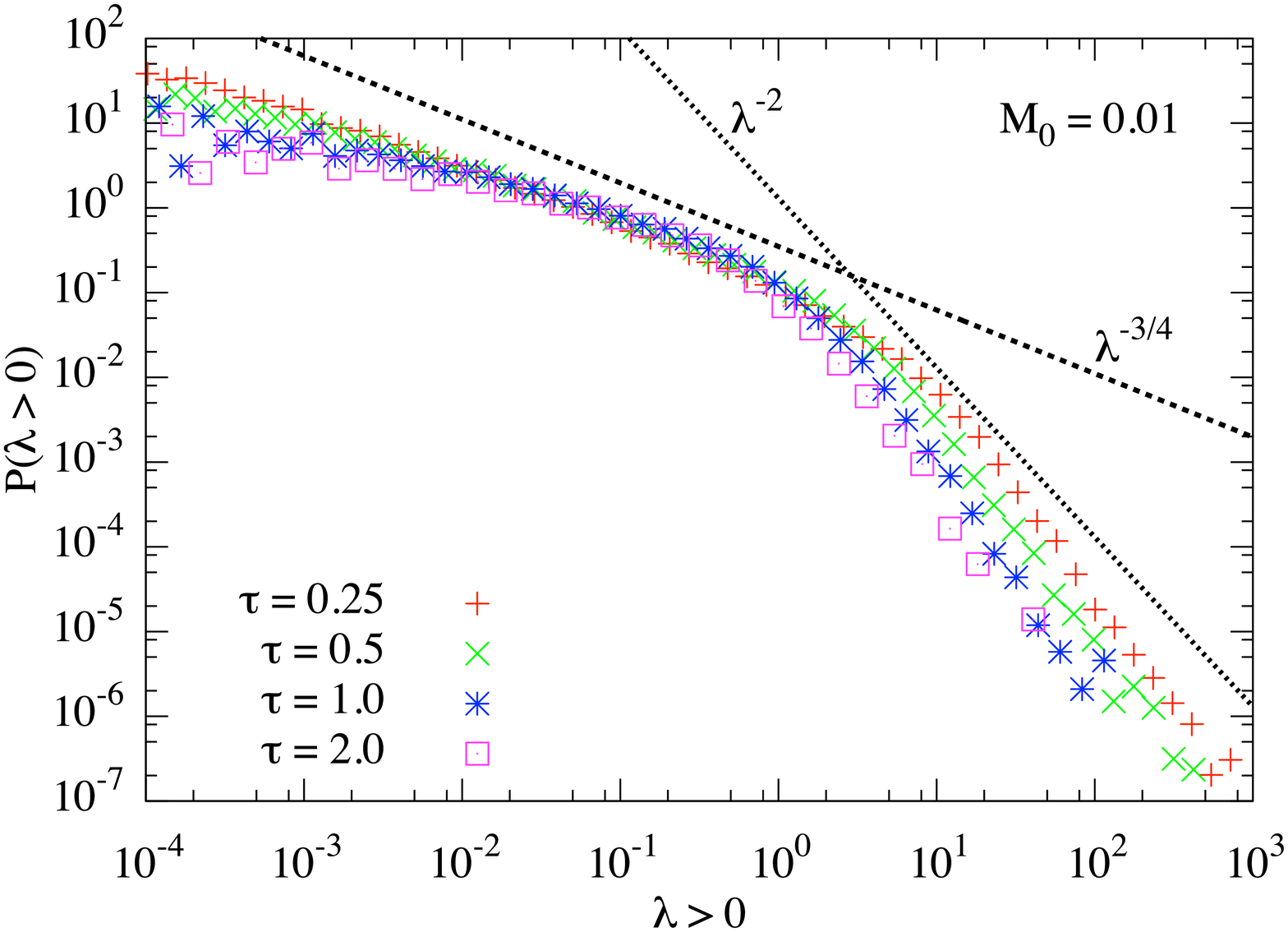}
\includegraphics[width=85mm]{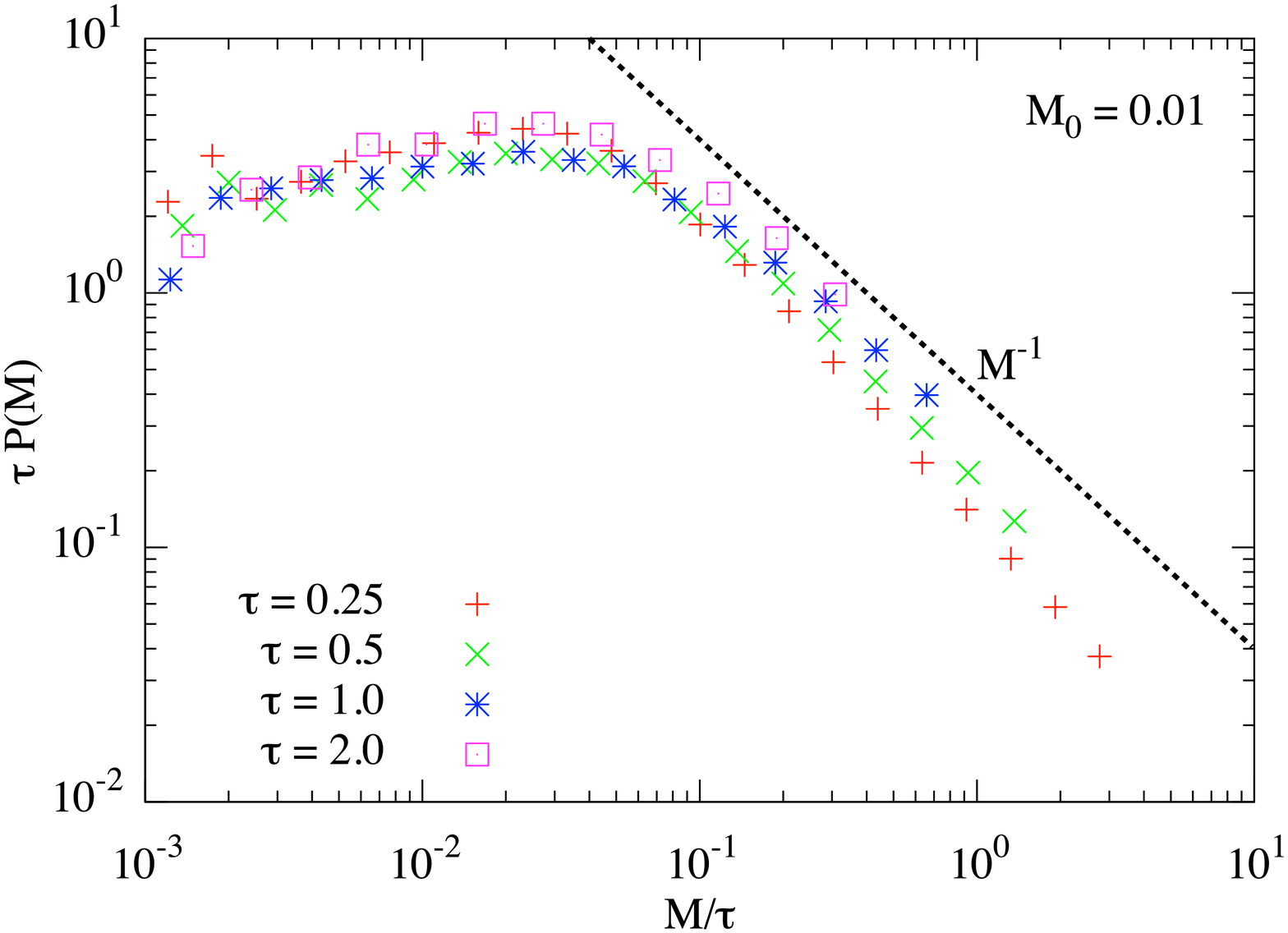}
\caption{{\it Left panel:} PDF of $\lambda=\tsigma M$ for $\lambda>0$ in 
Model II, \eqsdash{eq:M2}{eq:sigma2} with initial magnetic energy $M=0.01$. 
The PDF is shown at $\tt=0.25$, $0.5$, $1$ and $2$ and is approximately stationary. 
Slopes corresponding to $\lambda^{-3/4}$ and $\lambda^{-2}$ are shown for reference
(they are derived in \apref{ap:M2}). 
{\it Right panel:} PDFs of $M$ for the same simulation, shown for the same 
times, with $M$ rescaled by the time $\tt$ (the PDF evolves self-similarly; 
see \apref{ap:M2}). The slope corresponding to 
$M^{-1}$ is shown for reference. Note that realizations that 
reach $M=1$ are removed from our simulation, so this PDF is just for the 
remaining ones at any given time.} 
\label{fig:M2_pdfs}
\end{figure*}

As we explained in \secref{sec:M2qual}, the dynamo is nonlinear at all times 
and so, in some typical sense, $\dd_\tt M$ becomes a constant independent of $M$ (\eqref{eq:Msecular}). 
In the stochastic system given by \eqsdash{eq:M2}{eq:sigma2}, this behaviour 
can be teased out by formally introducing a new stochastic variable, $\lambda=\tsigma M$, so 
\eqref{eq:M2} becomes
\beq
\dd_\tt M = \lambda 
\label{eq:M_lambda}
\eeq
and \eqref{eq:sigma2} can be transformed accordingly into an equation for $\lambda$. 
It then turns out that $\lambda$ has a stationary distribution with most probability 
around $\lambda\sim1$ (as should be expected from the estimate $\tsigma\sim1/M$). 
This distribution is strongly intermittent:
the PDF of $\lambda$ has a power-law tail, $P(\lambda)\sim \lambda^{-2}$ 
at $\lambda\gg1$, and so a logarithmically divergent mean.
%(at $\lambda\ll1$, $P(\lambda)$ is also a power law; see \apref{ap:M2}). 
Since from \eqref{eq:M_lambda} it follows that $\la M\ra = \la\lambda\ra\tt$, the mean magnetic energy 
also diverges; indeed, its PDF trns out to have a power-law tail, $P(M)\sim M^{-1}$ at $\tt\ll M\ll 1$.
The PDFs of $\lambda$ and $M$ obtained by numerical solution 
of \eqsdash{eq:M2}{eq:sigma2} are shown in \figref{fig:M2_pdfs}; 
note that the PDF of $M$ evolves self-similarly in time.  

Obviously, the divergence of $\la M\ra$ does not mean that the magnetic energy 
is infinite or, worse still, that its PDF is non-normalizable as an $M^{-1}$ tail 
would imply. In fact, the distribution of $M$ is regularized at $M\sim1$ 
because for $M\gtrsim1$, the rate of strain $\tsigma$ 
no longer spends most of the time outside the stable interval $[-2M,M]$
and the dynamics start to look more akin to the conventional 
dynamo of \secref{sec:zeroD}. Furthermore --- and more to the point, physically, 
--- the field growth must saturate at $M\gtrsim1$ via a nonlinear mechanism 
not included in our model (in our numerical simulations, we simply remove the 
realizations that reach $M=1$). Physically, the divergence of $\la M\ra$ 
on time scales $\tt\sim1$ is a statistical expression of the fact that individual 
realizations can have periods of intense growth that take them to $M=1$ 
in very short times $\tt\sim\tt_0\ll1$, as discussed in \secref{sec:M2time}. 

\subsection{Constraints on the Seed Field}
\label{sec:M2_constraints}

We now itemize a set of constraints on the magnetic field in the same way as we 
did in \secref{sec:M1_constraints} for Model~I, 
with typical ICM time scales given by \exref{eq:ICM_sigma}--\exref{eq:ICM_Omega}\\ 

\noindent (i) Plasma must remain magnetized with respect to the increased 
effective collisionality: $\nueff\sim\sigma\beta\sim (\eps/p)\beta^2 \ll \Omega_i$ 
(see \eqsand{eq:nueff}{eq:sigmaest}). This gives
\beq
\frac{1}{\beta} \gg\lt(\frac{\eps}{p\Omega_1}\rt)^{2/5} \sim 10^{-7},
\ {\rm or}\ 
M \gg \frac{\nuii^{3/5}}{\sigma_0^{1/5}\Omega_1^{2/5}}\sim 10^{-5}, 
\eeq
a much more stringent constraint than \eqref{eq:magnetized}, which used the 
``bare'' Coulomb collision rate.\\

\noindent (ii) The same lower bound on the magnetic 
field is obtained (coincidentally) by demanding that the rate of strain remain slower 
than the peak growth rates \exref{eq:gfir} and \exref{eq:gmir} of the firehose and 
mirror instabilities. The bound is the same as for Model~I, \eqref{eq:gmir_bound}, 
because the maximum rate of strain affecting the magnetic field is 
the same in Model~II and in the version of Model~I proposed 
in \secref{sec:multiscale} (see \eqsand{eq:sigmaest}{eq:sigma_lB}) 
--- indeed, as discussed in \secref{sec:M2qual}, setting the effective 
collisionality to pin the pressure anisotropy at the marginal level is 
equivalent to requiring the magnetic field to be just dynamically 
significant at the corresponding ``viscous'' scale, the principle 
used to set that scale in \secref{sec:multiscale}.\\  

\noindent (iii) The above constraints are in fact overriden by an even 
stronger requirement that the effective collision rate (not just the rate 
of strain) be smaller than the mirror's growth rate: 
imposing $\nueff \sim (\eps/p)\beta^2 \ll \gmir\sim \Omega_1/\beta^{5/2}$ gives
\beq
\frac{1}{\beta} \gg\lt(\frac{\eps}{p\Omega_1}\rt)^{2/9} \sim 10^{-5},
\ {\rm or}\ 
M \gg \frac{\nuii^{5/7}}{\sigma_0^{3/7}\Omega_1^{2/7}}\sim 10^{-3}. 
\label{eq:M2coll_bound}
\eeq 
Thus, as with Model~I, we again conclude that a relatively simple closure 
scheme represented by Model~II is not really adequate for 
describing the evolution of truly weak magnetic fields. 
We will discuss this issue further in \secref{sec:maggen}. 

\subsection{Summary for Model~II} 
\label{sec:M2_sum}

In this model, the pressure anisotropy is kept marginal by the enhancement 
of the effective collisionality of the plasma, owing to alleged scattering 
of particles off the microscale magnetic fluctuations caused by the pressure-anisotropy 
driven instabilities. This results in a dramatic decrease of the local 
viscosity of the plasma wherever and whenever the field is weak and changing. 
The turbulent cascade in such places can extend to very small scales,  
so the rate of strain becomes very large and very 
fast field growth results. Dynamical saturation can be achieved, at least by 
some fraction of realizations, on timescales that are smaller for weaker seeds 
than for strong ones (\eqref{eq:t2_fast}) --- and by the majority of realizations 
after one large-scale turnover time (\eqref{eq:t2}). Both the field and its 
rate of growth are extremely intermittent, with shallow power-law distributions 
(\secref{sec:M2stoch}), as one might intuitively expect in a physical system where 
the local value of viscosity depended on the local instantaneous magnetic field and 
its stretching rate. The dynamo is always nonlinear in the sense that
the magnetic-energy density is comparable to the kinetic-energy density 
of the motions that are dominantly amplifying it at any given moment in time 
(\secref{sec:M2qual}).

Thus, the dynamo in Model~II is fast (much faster than the regular MHD dynamo)\footnote{See 
discussion in \secref{sec:modelling} on why this is not seen in current 
numerical studies employing the enhanced-collisionality closure.} and very efficient.
However, in order for the model to be valid, scale 
separation is required between the smallest spatial and time scales of 
the fluid motions benefitting from diminished viscosity and the 
spatial and time scales of the mirror and 
firehose fluctuations that are making this diminished viscosity possible.
This requirement constrains the field (under the most pessimistic 
estimate, \eqref{eq:M2coll_bound}) to values just over an order of magnitude 
below the observed strength, rather like in Model~I. 
Thus, here again, the model of field evolution provided by the 
enhanced-collisionality closure is more suitable for studying the 
magnetized dynamics of the current state of the ICM than the genesis 
of the observed field from a tiny primordial seed (see further discussion 
in \secref{sec:maggen}).

\section{Discussion}
\label{sec:disc}

It is perhaps fair to say that the foregoing represents more an account of the 
state of our ignorance about the nature of plasma dynamo (or, more generally, 
the dynamics of magnetic field in weakly collisional envirinments) than 
a definitive solution even to a subset of the problem. Nevertheless, 
we believe it was useful to compose this catolgue of seemingly sensible modelling 
choices, {\em a prirori} limitations of their validity 
(see \secsand{sec:M1_constraints}{sec:M2_constraints}), and their possible 
consequences --- not least because moving these from 
the category of unknown unknowns to that of the known ones 
allows for a more informed discussion of the problems at hand. 

What are these problems that motivate the need to model the ICM at all? 
Here we focus on three of them; for the specific 
conclusions from our investigation of Models~I and II, we refer the reader 
to the summaries provided in \secsand{sec:M1_sum}{sec:M2_sum}. 

\subsection{Cosmic Magnetogenesis and Collisionless Plasma Dynamo}
\label{sec:maggen}

It is a long-standing question how the magnetic field in the Universe 
has managed to grow from a small primordial seed to the observed 
value (in the $\mu$G range in galaxy clusters). This requires an 
amplification by at least three and possibly many more orders of 
magnitude over a period of a few billions of years (the estimates 
for the primordial field vary from $10^{-21}$ to, very optimistically, 
$10^{-9}$~G; see \citealt{Durrer13}). The plasma is magnetized 
and weakly collisional (dilute) starting for $B\gtrsim10^{-17}$~G, so 
conventional MHD description is unsuitable and a model of field 
evolution incorporating pressure-anisotropy-driven microinstabilities is required. 

Normally one assumes that there would be a healthy scale separation 
between the chaotic motions of the ICM that amplify 
the field and the firehose and mirror instabilities that 
limit the pressure anisotropy arising from any attempt 
to change $B$. The two models of how they do that considered 
above were concerned with the realtionship between the pressure 
anisotropy and the rate of change of the field while taking the 
action of the instabilities to be instantaneous. 
Two very different field-evolution scenarios have emerged 
(slow for Model I, fast for Model II) --- but what the models 
have in common is the dramatic reduction in the field-parallel 
viscous stress (which is equal to the pressure anisotropy). 
As a consequence, the turbulent rates of strain responsible 
for amplifying the field are expected to be much larger and to occur 
at much smaller scales than is usually assumed for the ICM. 
Keeping these scales separated from the scales (in time and space) 
at which the firehose and mirror fluctuations occur 
is only possible for magnetic fields merely one or two orders of 
magnitude below the observed level. Thus, in order to understand how 
the field can grow from a primordial seed to values of order 
$10^{-8}-10^{-7}$~G, we need a theory of a {\em fully collisionless dynamo 
operating with no scale separation between plasma flows and 
pressure-anisotropy-driven instabilities}. Is there such a dynamo? 
How fast is it? These questions are open --- and they cannot be answered 
by solving fluid equations with any microphysical closure that assumes 
an instantaneous adjustment of the pressure anisotropy to marginal level, 
but rather require a fully kinetic treatment. This is a hard problem, 
but the good news is that at least its numerical solution appears less 
challenging if no scale separation between the motions and the 
instabilities is required. 

%\red{Low seed field: loss of scale separation, collisionless dynamo, 
%including instabilities, possibly secondaries etc. A bit like 
%Weibel or Bell mechanisms? \citet{Schoeffler13}. 
%As $B$ grows, the more conventional dynamo rises from this mess?}

\subsection{Understanding Observed Magnetic Field and ICM Motions}

As the field grows closer to the observed level, the instabilities do 
become instantaneous and so the dynamics of the field and the plasma 
can perhaps be described by MHD equations with a microphysical closure 
represented by one of the two models we have studied. Understanding the 
structure of the saturated field --- and of the turbulence into which it 
is embedded --- is a fascinating problem, also quite open. In particular, 
it is entirely unclear what sets the spatial scale of the observed 
magnetic field. Arguably this is actually a more interesting problem than 
the magnetogenesis as the field appears to be of dynamically important 
strength everywhere it has been measured, so theories of how it got there 
are not observationally falsifiable (except perhaps in the laboratory; 
see, e.g., \citealt{Spence09}). In contrast, turbulence measurements in the ICM 
over a range of scales are a growing industry \citep[e.g.,][]{Schuecker04,Vogt05,Churazov12,Sanders13},
so a good understanding of its magnetofluid dynamics is quite indispensable
--- and not possible without a model of how magnetic field can change in a 
moving weakly collisional plasma. 
This point applies with even greater force to the multitudinous observations and 
attendant modelling of various large- and medium-scale ordered motions 
in the ICM, which invariably require dragging magnetic field around 
\citep[e.g., rising bubbles; see][and references therein]{Churazov13}.  

%\red{NB: equipartitioned field because res.\ scale is close to visc.?
%Cutoff scale is instability scale? For either model.} 

\subsection{Heating of the ICM}

An interesting and important non-trivial consequence of the closure one assumes 
for the pressure anisotropy and for the evolution of the magnetic field 
is the thermal stability of the ICM and hence 
the existence or absence of the so-called cooling catastrophe 
\citep[see][and references therein]{Kunz11}. 
The viscous heating 
rate per unit volume in a magnetized, subsonically moving plasma is 
\beq
Q_{\rm visc} = (\pperp-\ppar)\,\gamma
\sim \nuii p \Delta^2 \sim \frac{\nuii p}{\beta^2},
\label{eq:Qvisc}
\eeq 
where the last expression follows by assuming that the pressure anisotropy 
is marginal with respect to firehose or mirror instability conditions 
\citep[cf.][]{Lyutikov07}. 
Thus, even though the viscous heating comes from the dissipation 
of plasma motions, the final expression for it does not appear to 
be related to them except via the local value of $\beta$ (which, one assumes, 
is set by the saturated level of the magnetic field and, therefore, 
related to the kinetic-energy density of the turbulence). 

Under the assumptions of Model~I, the collision rate in \eqref{eq:Qvisc} 
is the Coulomb collision rate and so we have a specific local relationship between 
the heating and the local values of magnetic field $B$, density $n$ and temperature $T$ 
of the ICM. Balancing $Q_{\rm visc}$ with the radiative cooling rate of the ICM, 
which depends on $n$ and $T$, produces a definite relationship between these 
parameters and the magnetic field and leads to a thermally {\em stable} equilibrium 
with reasonable values of $Q_{\rm visc}$ and $B$ for typical cluster-core 
conditions \citep{Kunz11}. 
%One assumes in this argument that only the viscous heating matters 
%(i.e., that the Alfv\'enic part of the turbulence can be ignored) and that the energy 
%deposition into the ICM turbulence from various sources (AGNs, galaxy wakes etc.) will 
%always adjust to the maximum locally allowed dissipation rate set by \eqref{eq:Qvisc} 
%(which can itself adjust via the local magnetic-energy density and, therefore, the local 
%energy density of the turbulent motions). 

Turning to Model~II, we must replace in \eqref{eq:Qvisc} 
$\nuii\to\nueff\sim\sigma\beta$. Thus, the collisionality is no longer set 
by the equilibrium state of the plasma but adjusts to the local rate of 
strain. Using \eqref{eq:sigmaest} and $\sigma_0\sim(\eps\nuii/p)^{1/2}$, 
we find simply that $\nueff\sim\eps\beta^2/p$ and 
so\footnote{In Model~I, this equation also holds 
but is non-trivial and allows one to determine the injection scale 
of the turbulence \citep{Kunz11}, whereas in Model~II, it is automatic 
and provides no further information.} 
\beq
Q_{\rm visc} \sim \eps. 
\eeq
Thus, whatever energy is injected into turbulence will be dissipated 
by heating, with no more microphysical constraints on the heating 
rate. This is a situation more familiar to those used 
to dealing with standard fluid turbulence problems, although here it arises 
not because the scale of the motions adjusts to accommodate a given dissipation rate 
but because the collisionality of the plasma does. This means that whatever determines 
the thermal stability of the ICM in this scenario has to do with how 
large-scale energy is deposited into the ICM turbulence 
--- an outcome that may be conceptually satisfying to anyone who 
feels that microphysics should never matter (although it does 
matter for deciding whether this scenario is in fact correct).   

\subsection{A Comment on Existing Numerical Simulations with Microphysical Closures}
\label{sec:modelling}

We are not aware of any studies so far to simulate the ICM under the assumptions of Model~I. 
The effective enhancement of the collision rate that underpins Model~II 
is more straightforward to implement in MHD-CGL equations 
and this has been done by several groups \citep{Sharma06,Meng12,Kunz12,Lima14}. 
Relatively little qualitative difference with the standard collisional MHD case 
was found, suggesting that the effect of plasma microphysics is simply to 
render collisionless plasma effectively collisional \citep{Lima14}. 
This is a tempting conclusion, which, if true, would relieve the ICM 
modelling community of a serious headache.  
Putting aside the question of how likely the firehose 
and especially mirror instabilities in fact are to cause enhanced particle scattering 
(see \secref{sec:status}), if they do, this still appears to produce 
a highly complex situation with spatially and temporally intermittent 
local viscosity and hence very different field evolution than in collisional MHD
(\secref{sec:M2}). However, one can only capture this complexity in a numerical simulation 
if it is the intermittent local viscosity and not numerical grid dissipation or some 
other form of fixed-scale regularization that determines the cutoff scale 
for the turbulence. If, on the other hand, a fixed-scale regularization 
(effectively, a small isotropic viscosity similar to the one present in 
the collisional MHD) is present and, given limited resolution, 
overrides the plasma viscosity (much diminished owing to enhanced collisionality), 
the effect of the microphysical closure 
is simply to disconnect the pressure anisotropy from the stretching rate 
and, therefore, from having any influence on the evolution of the 
magnetic field --- either directly or via its effect on the local viscosity 
of the plasma. Note that the pressure anisotropy will still have a role 
in modifying the tension force --- an order-unity enhancement at the mirror 
threshold (a prefactor of $3/2$ when $\Delta=1/\beta$; see \eqref{eq:u}) 
and a full suppression at the firehose threshold ($\Delta=-2/\beta$). 
The latter effect appears potentially to be the more important, but in turbulent 
and dynamo situations, the mirror-unstable regions tend to dominate 
\citep{Sharma06,Lima14}.\footnote{Note, however, the crucial role that 
the firehose threshold appears to play in magnetic reconnection 
\citep{Schoeffler11,Matteini13}.
If reconnection of the magnetic fields must be understood quantitatively 
in order to predict the structure of the saturated dynamo states, 
the firehose regions may turn out to be more important than their 
spatial sparseness might suggest.}\\  

To conclude, the results presented above highlight the extent to which weakly collisional 
magnetized plasmas have the potential to surprise us and the importance of getting to grips 
with the rich microphysical world that, while unobservable directly 
(except perhaps in the solar wind), undelies all 
observable large-scale dynamics and thermodynamics of these plasmas.

\section*{Acknowledgments}
It is a pleasure to thank M.~Kunz, S.~Cowley, G.~Hammett, D.~McHardy, E.~Quataert, F.~Rincon 
and D.~Sokoloff for useful discussions and comments on aspects of this work. 
The work of F.~M.\ was carried under the ENS, Paris undergraduate research internship scheme. 

\appendix

\section{Zero-Dimensional Dynamo}
\label{ap:zeroD}

Here we provide the full solution of the ``zero-dimensional dynamo'' 
introduced in \secref{sec:zeroD}. The calculation 
is standard but, as far as we know, is not readily available in textbooks 
in this form. 

We wish to consider the following equations
\bea
\label{eq:Bap}
\dd_t B &=& \sigma B,\\
\dd_t \sigma &=& - \tau_c^{-1}\sigma + \sqrt{2D}\,\chi(t),
\label{eq:Langevin}
\eea
where in the Langevin equation \exref{eq:Langevin}, 
$\tau_c$ is the correlation time and $D$ the diffusion coefficient
(\eqref{eq:sigma0} has $\tau_c=1/2\sigma_0$ and $D = 2\sigma_0^3$). 

\subsection{Fokker--Planck Equation}

The joint time-dependent probability-density function (PDF) of $\sigma$ and $B$ is 
$P(t,\sigma,B) = \la\tP\ra$, where 
$\tP = \delta(\sigma - \sigma(t))\delta(B - B(t))$, 
where $\sigma$ and $B$ are random variables, whereas $B(t)$ and $\sigma(t)$ 
are solutions of \eqsand{eq:Bap}{eq:Langevin}. Then
\bea
\nonumber
\dd_t\tP &=& - \dd_\sigma\tP \dd_t \sigma(t) + \dd_B\tP \dd_t B(t) \\
&=& -\dd_\sigma \lt(- \tau_c^{-1}\sigma + \sqrt{2D}\,\chi\rt)\tP 
- \dd_B \sigma B \tP.
\label{eq:tP}
\eea
Averaging this, we get
\beq
\dd_t P = - \sqrt{2D}\,\dd_\sigma\la\chi\tP\ra + \tau_c^{-1}\dd_\sigma\sigma P 
- \sigma\dd_B BP.
\label{eq:P}
\eeq
The average can be calculated by formally integrating \eqref{eq:tP}, 
\beq
\tP(t) = \!\int^t\!\!\rmd t'
\lt[\dd_\sigma\lt(\tau_c^{-1}\sigma - \sqrt{2D}\,\chi(t')\rt)
- \dd_B \sigma B\rt]\!\tP(t'),
\eeq
and using the fact that $\chi(t)$ and $\tP(t')$ are uncorrelated for $t\ge t'$ 
($\tP$ only depends on the past values of $\chi$, but not the present or the 
future), $\la\chi\ra=0$ and $\la\chi(t)\chi(t')\ra = \delta(t-t')$: 
\beq
\la\chi(t)\tP(t)\ra = -\frac{1}{2}\sqrt{2D}\,\dd_\sigma P.
\eeq
Combined with \eqref{eq:P}, this gives the Fokker--Planck equation for the 
joint PDF: 
\beq
\dd_t P = D\, \dd_\sigma^2 P + \tau_c^{-1}\dd_\sigma\sigma P 
- \sigma\dd_B BP.
\label{eq:FP}
\eeq 

\subsection{Moments of $B$}

Now let $P_n(\sigma) = \int_0^\infty\rmd B B^n P(\sigma,B)$. Then $P_0(\sigma)$ 
is the PDF of $\sigma$ and $\int\rmd\sigma P_n(\sigma) = \la B^n\ra$ 
are the moments of $B$. $P_n$ satisfies
\beq
\dd_t P_n = D\, \dd_\sigma^2 P_n + \tau_c^{-1}\dd_\sigma\sigma P_n + n\sigma P_n.
\label{eq:Pn}
\eeq
If we look for solutions in the form
\beq
P_n = \psi_n(x)\exp\lt(\gamma_n t - \frac{\sigma^2}{4D\tau_c}\rt),\quad
x = \frac{\sigma - 2D\tau_c^2 n}{\sqrt{2D\tau_c}},
\label{eq:ansatz}
\eeq
then $\psi(x)$ satisfies
\beq
\psi_n'' - x^2\psi_n = -(1 - 2\gamma_n\tau_c + 2D\tau_c^3 n^2)\psi_n. 
\eeq
This is a Schr\"odinger equation for a harmonic oscillator. 
The non-oscillating solution is the ground state, corresponding to the 
expression in the parentheses on the right-hand side (the energy) being equal to~$1$. 
Then $\gamma_n = D\tau_c^2 n^2$ and $\psi_n = C_n e^{-x^2/2}$, where $C_n$ are constants. 
Assembling this with \eqref{eq:ansatz}, we get 
\beq
P_n = \frac{\tC_n}{\sqrt{2\pi D\tau_c}}
\exp\lt[D\tau_c^2 n^2 t - \frac{(\sigma - D\tau_c^2n)^2}{2D\tau_c}\rt],
\label{eq:slnPn}
\eeq
where some $\sigma$- and $t$-independent factors have been absorbed into the 
new constant $\tC_n$. 
Note that for $n=0$ and $\tC_0=1$, this gives a Gaussian distribution for $\sigma$, 
with $\la\sigma^2\ra = D\tau_c$. For $\tau_c = 1/2\sigma_0$ and $D = 2\sigma_0^3$, 
we get $\la\sigma^2\ra = \sigma_0^2$, as stated in \secref{sec:zeroD}. 

Integrating \eqref{eq:slnPn} over $\sigma$, we obtain 
the time evolution of the moments of the magnetic field:
\beq
\la B^n\ra = \tC_n e^{D\tau_c^2 n^2 t}. 
\label{eq:Bn}
\eeq
For $\tau_c = 1/2\sigma_0$ and $D = 2\sigma_0^3$, this becomes the expression 
quoted in \secref{sec:zeroD}. We see that the constants $\tC_n = \la B_0^n\ra$ 
are the moments of the initial distribution of $B$. 
For simplicity, we may assume that the field starts 
with the same value $B_0$ in all realizations and that $B$ is normalized to that value; 
then all $\tC_n=1$.  

\subsection{PDF of $B$}

The quickest way to calculate the PDF of $B$ 
is to notice that there was nothing in the calculation above that required 
$n$ to be discrete or even real. Therefore, letting $n=i\lambda$ in \eqref{eq:Bn}, 
find 
\beq
\la e^{i\lambda\ln B}\ra =  e^{-D\tau_c^2 \lambda^2 t}.
\eeq
The left-hand side is the characteristic function (the Fourier transform) of the 
PDF of $\ln B$. Inverse-Fourier transforming in $\lambda$ and expressing
the result as the PDF of $B$, we get
\beq
P(B) = \frac{1}{B\sqrt{4\pi D\tau_c^2 t}}\,\exp\lt(-\frac{\ln^2B}{4 D\tau_c^2 t}\rt),
\label{eq:PB}
\eeq
the lognormal distribution quoted in \secref{sec:zeroD} (with $4D\tau_c^2 = 2\sigma_0$). 

Note that a typical realization grows subexponentially: 
$\ln B \sim 2\tau_c\sqrt{Dt}$ even though all the moments $\la B^n\ra$
grow exponentially, a property that 
implies that a very large number of realizations must be used in a numerical solution 
in order to capture the correct intermittent dynamo behaviour over any given time $t$ 
\citep[cf.][]{Art06}. This is, however, a feature to some extent particular 
to the zero-dimensional model: in 3D, a typical growing realization does grow exponentially 
(see discussion and references at the beginning of \secref{sec:meanrate}), although 
its growth rate is still different from the growth rate of the magnetic energy 
and the point about the necessity of good statistics stands. 

\subsection{Time to Saturation}
\label{ap:tau}

In terms of the dimensionless quantities used in most of this paper, 
$M=\nuii/\sigma_0\beta_0$ and $\tau=2\sigma_0t$, the PDF of the magnetic 
energy is 
\beq
P(\tt,M|M_0) = \frac{1}{M\sqrt{4\pi\tt}}\,\exp\lt[-\frac{\ln^2(M/M_0)}{4\tt}\rt],
\eeq
which is also the probability of reaching magnetic energy $M$ at time $\tau$ 
starting with $M_0$ at $\tau=0$. 
Given $P(\tt,M|M_0)$, one can calculate the probability $P_1(\tt|M_0)$ of reaching 
the dynamical level $M=1$ (\eqref{eq:sat}) 
for the first time at time $\tt$ (the ``first-passage time'') 
via the standard relation 
\beq
\label{eq:P1}
P(\tt,1|M_0) = \int_0^\tt\rmd\tt' P_1(\tt'|M_0) P(\tt-\tt',1|1).
\eeq
Inverting the integral operator in the right-hand side via the Laplace transform, we get 
\beq
P_1(\tt|M_0) = \frac{|\ln M_0|}{2\sqrt{\pi}\tt^{3/2}}\exp\lt(-\frac{\ln^2 M_0}{4\tt}\rt).
\label{eq:P1_0D}
\eeq
This immediately allows us to calculate the fraction of the realizations that eventually reach $M=1$:
\beq
\fmax=\int_0^\infty\rmd\tt P_1(\tt|M_0) = 1
\label{eq:fmax_0D}
\eeq 
(i.e., they all do in this model). Note that the mean time for this to happen, 
\beq 
\la\tt\ra = \frac{1}{\fmax}\int_0^\infty\rmd\tt\tt P_1(\tt|M_0),
\eeq
diverges because $P_1\sim \tt^{-3/2}$ at $\tt\gg1$
(so the 100\% ``dynamo efficiency'' implied by \eqref{eq:fmax_0D} takes a long 
time to consummate). The characteristic time for a typical realization to reach 
dynamically relevant fields can be read off from the exponential factor 
in \eqref{eq:P1_0D}: $\tt\sim\ln^2 M_0$. 
However, the exponential growth of the mean magnetic energy (\eqref{eq:Bn}) 
implies that the realizations that dominantly contribute to 
$\la M\ra$ (and all other moments) only require $\tt \sim \ln M_0$ 
(hence \eqref{eq:t0}).\footnote{This highlights the point that what is meant 
by a ``dynamo'' and how fast that dynamo is considered to be is to an extent 
a matter of definition: do we wish the mean magnetic energy $\la B^2\ra$ to grow 
exponentially? do we wish a typical realization to do so? In the rather simplistic 
zero-dimensional model adopted here, all realizations do eventually reach the nonlinear 
threshold \exref{eq:sat}, albeit at sub-exponential rates (see \apref{ap:tau}), 
but in a standard model of a more realistic 3D MHD dynamo taking into account also 
the effect of resistivity, most realizations in fact decay superexponentially 
while both the typical growing realizations and 
the magnetic energy grow exponentially, albeit at different rates 
\citep{Zeldovich84,Chertkov99,Sch04,MHDbook}. 
In our treatment of Model~I, we encounter a case of very difficult 
field growth (\secref{sec:onescale}).} 

\section{Magnetic Field Statistics and Growth Times in Model I with One-Scale Flow}
\label{ap:M1}

Here we describe a very simple way to understand the behaviour of the one-scale 
version of Model I (\secsand{sec:onescale}{sec:meanrate}). 
Let $M_0\le M\ll 1$. Since $\tsigma$ is order unity (\eqref{eq:sigma1}), 
the rate of strain will spend most of the time outside the stable interval $\lt[-2M,M\rt]$ 
and so, as $\tsigma$ fluctuates between positive and negative values, the 
effective stretching rate will be alternately pinned at the mirror or firehose 
threshold. Thus, we may approximately replace \eqref{eq:M1} with 
\beq
\dd_\tt M = \lt\{
\begin{array}{cl}
M^2, & \tsigma > 0,\\
-2M^2, & \tsigma < 0.
\end{array}
\rt.
\label{eq:Mdiscrete}
\eeq
Furthermore, let us take the long-time limit, $\tt\gg1$, treat 
$\tau$ as a discrete counter with step size $\Dtt\sim1$ 
and $\tsigma(\tau)$ as a sequence of discrete independent trials with either 
positive or negative outcome, each with probability $p=1/2$ (this is reasonable 
because the correlation time of $\tsigma$ is order unity, but the crude 
nature of the model will leave us with the need to fit the numerically obtained 
distribution to the analytical result by choosing a suitable value 
of $\Dtt$, which will indeed turn out to be of order unity; see \apref{ap:M1_PM}). 
Integrating \eqref{eq:Mdiscrete}, we get 
\beq
\frac{1}{M_0} - \frac{1}{M(\tt)} = \Dtt\sum_{i=1}^{\tt/\Dtt} x_i,\quad
x_i = \lt\{
\begin{array}{cl}
1, & p = 1/2,\\
-2, & p = 1/2.
\end{array}
\rt.
\label{eq:xi}
\eeq

\begin{figure}
\includegraphics[width=80mm]{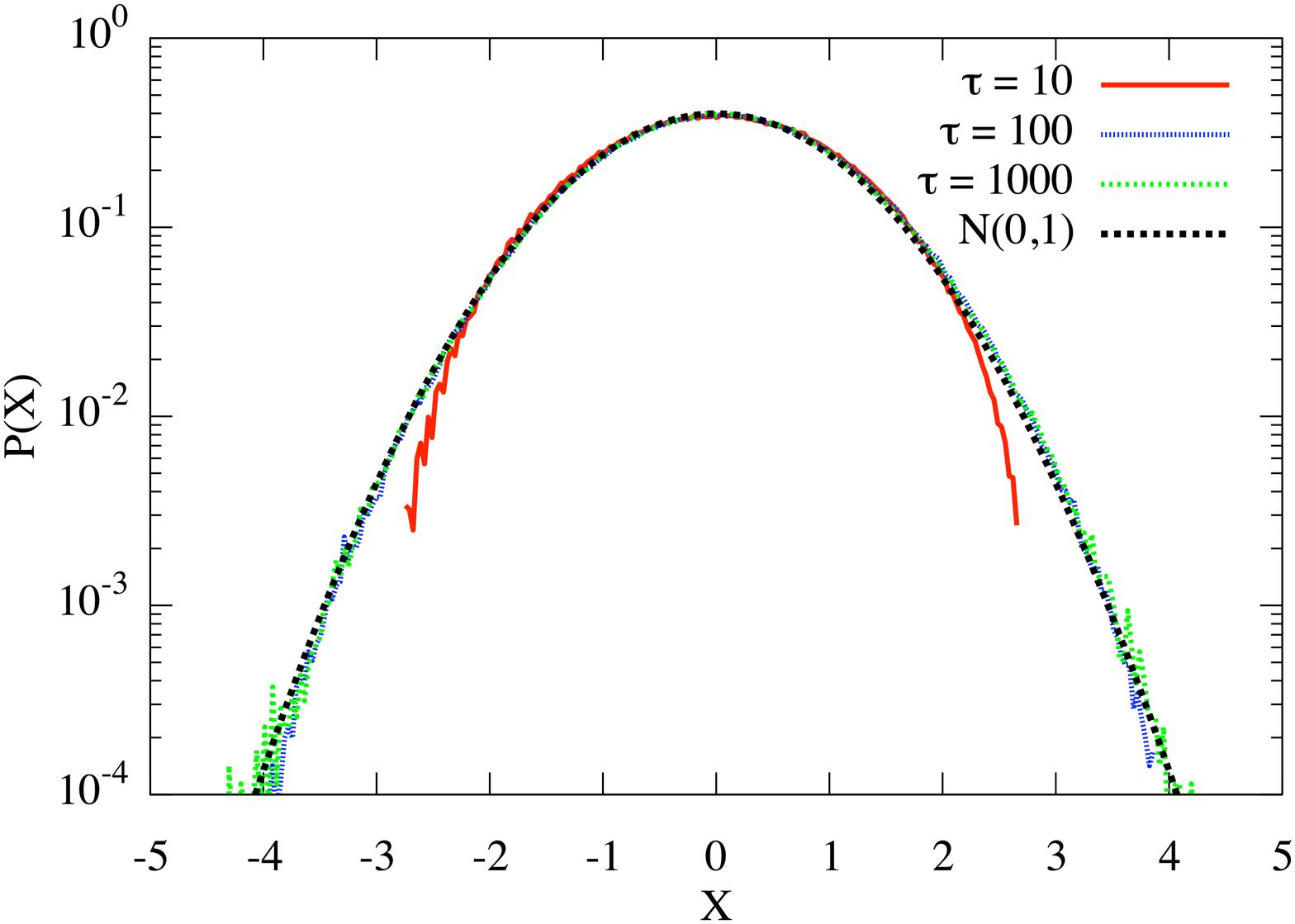}
\caption{PDF of the quantity $X$ given by \eqref{eq:X}. 
Several times are shown for the case with $M_0=0.05$, together 
with the unit normal PDF \exref{eq:PX}. The best fit is obtained 
for $\Dtt=1.4$.}
\label{fig:M1_pdfM} 
\end{figure}

\subsection{PDF of $M$}
\label{ap:M1_PM}

The mean $\la x_i\ra = -1/2$ and the variance $(\la x_i^2\ra - \la x_i\ra^2)^{1/2} = 3/2$, 
so, by Central Limit Theorem, the quantity 
\beq
X = \frac{2}{3}\sqrt{\frac{\tt}{\Dtt}}
\lt[\frac{1}{\tt}\lt(\frac{1}{M_0} - \frac{1}{M}\rt) + \frac{1}{2}\rt]
\label{eq:X}
\eeq
has the unit normal distribution 
\beq
P(X) = \frac{1}{\sqrt{2\pi}}\,e^{-X^2/2}. 
\label{eq:PX}
\eeq
The PDF of $X$ found in the numerical solution of \eqsand{eq:M1}{eq:sigma1} is shown 
in \figref{fig:M1_pdfM}; we find that $\Dtt=1.4$ gives the best fit. 
Thus, the PDF of the magnetic energy at time $\tt$, having started 
with $M_0$ at time $0$, is
\bea
\nonumber
&&\!\!\!\!\! P(\tau,M|M_0) =\\ 
&&\quad
\frac{M^{-2}}{\sqrt{9\pi\tau\Dtt/2}}
\exp\lt[-\frac{2}{9\tt\Dtt}\lt(\frac{1}{M_0} - \frac{1}{M} + \frac{\tt}{2}\rt)^2\rt].
\label{eq:PM_M1}
\eea

To understand the time evolution of this distribution, note that,  
from \eqref{eq:X}, 
\beq
M = \frac{1}{1/M_0 + \tt/2 - 3X\sqrt{\tt\Dtt}/2}.
\label{eq:MvsX}
\eeq
We see that not much happens for $\tt\ll 1/M_0$; for $\tt\gg1/M_0$,  
the typical realizations decay secularly with time. The growing ones are due 
to the events with $X\gtrsim\sqrt{\tt}$, which are increasingly rare 
as time goes on (the mean tendency for $M$ to decay due to the 
asymmetry between the firehose and mirror thresholds wins in the long run). 
The explosively growing realizations are typically those with 
$X\sim \sqrt{\ttdyn} \sim 1/\sqrt{M_0}$ 
and their fraction is $f \sim \int^\infty_{\sqrt{\ttdyn}} \rmd X P(X)\sim \exp(-1/M_0)$, 
in line with \eqref{eq:f1}. 

Finally, letting $\tt\gg1$ and averaging powers of $M$ using \eqref{eq:MvsX} 
and the distribution \exref{eq:PX}, we find that all moments of the magnetic 
energy decay with time: 
\beq
\la M^n\ra \approx \frac{1}{\lt(1/M_0 + \tt/2\rt)^n} \to 0. 
\eeq
 
\subsection{Time to Saturation}
\label{ap:M1_tau}

Since we know the $M_0\to M$ transition probability \exref{eq:PM_M1}, we can calculate 
the probability to reach $M=1$ in time $\tt$ in the way described in \apref{ap:tau} 
(via \eqref{eq:P1}). The result is 
\beq
P_1(\tt|M_0) = \frac{\lt|\ttdyn\rt|}{\sqrt{9\pi\Dtt\tt^3/2}}
\exp\lt[-\frac{2}{9\Dtt\tt}\lt(\ttdyn + \frac{\tt}{2}\rt)^2\rt],
\label{eq:P1_M1}
\eeq 
where $\ttdyn=1/M_0 - 1$. 
Therefore, the fraction of realizations that ever make it to $M=1$ is 
(assuming $M_0<1$)
\beq
\fmax = \exp\lt[-\frac{4}{9\Dtt}\lt(\frac{1}{M_0} - 1\rt)\rt] 
\label{eq:fmax_M1}
\eeq
and the mean time for them to do it is 
\beq
\la\tt\ra = 2\lt|\frac{1}{M_0} - 1\rt| = 2|\ttdyn|.
\eeq
These are the quantitative versions of the estimates \exref{eq:f1} and \exref{eq:t1}, 
respectively. Note that \eqref{eq:P1_M1} implies that at long times, $\tt\gg\ttdyn$, 
the PDF of the time to $M=1$ is $\propto \tt^{-3/2}\exp(-\tt/18\Dtt)$ 
(see the right panel of \figref{fig:M1_tau}). 

\subsection{Case of Non-Zero Mean Stretching Rate}
\label{ap:meanrate}

The above calculations are easily generalized to the case of non-zero mean stretching rate 
(\secref{sec:meanrate}), which amounts to letting $x_i=1$ with probability $p>1/2$ 
and $x_i=-2$ with probability $1-p$ in \eqref{eq:xi}. 
Then $\la x_i\ra = 3p-2$, $(\la x_i^2\ra - \la x_i\ra^2)^{1/2} = 9p(1-p)$ 
and so \eqref{eq:X} becomes 
\beq
X = \sqrt{\frac{\tt}{9p(1-p)\Dtt}}
\lt[\frac{1}{\tt}\lt(\frac{1}{M_0} - \frac{1}{M}\rt) - (3p-2)\rt],
\label{eq:X_p}
\eeq
which is distributed normally (\eqref{eq:PX}). Therefore,
\beq
M = \frac{1}{1/M_0 - (3p-2)\tt - 3X\sqrt{p(1-p)\tt\Dtt}/2}.
\label{eq:MvsX_p}
\eeq
If $p>2/3$, this explodes at $\tt=1/(3p-2)M_0$. Accordingly, assuming $\tt\gg1$, 
we find that all moments of $M$ explode: 
\beq
\la M^n\ra \approx \frac{1}{\lt[1/M_0 - (3p-2)\tt\rt]^n}.
\eeq

The generalized version of \eqref{eq:P1_M1} for the probability of reaching 
$M=1$ at time $\tt$ is 
\bea
\nonumber
&&\!\!\!\!\!P_1(\tt|M_0) =\\ 
&&\frac{\lt|\ttdyn\rt|}{\sqrt{18\pi p(1-p)\Dtt\tt^3}}
\exp\lt\{-\frac{\lt[\ttdyn - (3p-2)\tt\rt]^2}{18p(1-p)\Dtt\tt}\rt\}.
\eea 
The fraction of the realizations that make it is the integral of the above:
\bea
\fmax &=& \exp\lt[-\frac{2(2-3p)}{9p(1-p)\Dtt}\lt(\frac{1}{M_0} - 1\rt)\rt],\quad p<\frac{2}{3},\\ 
\fmax &=& 1, \quad p\ge\frac{2}{3} 
\eea
and $\la\tt\ra = |\ttdyn|/|3p-2|$. Thus, for $p>2/3$, all realizations reach $M=1$ 
on the characteristic timescale $\sim\ttdyn$.

\section{Stochastic Nonlinear Plasma Dynamo in Model II}
\label{ap:M2}

Here we derive some analytical results for the 
dynamo model given by \eqsdash{eq:M2}{eq:sigma2}, to support the 
qualitative summary in \secref{sec:M2stoch}. 

Consider \eqsand{eq:M2}{eq:sigma2}. 
The expectation from the qualitative discussion in \secref{sec:M2qual} 
is that the magnetic energy will, in some typical sense, grow linearly 
in time at a rate that is of order unity in rescaled variables, $M\sim\tt$ 
(\eqref{eq:Msecular}). In a stochastic system, this will be a random 
process, so, anticipating the form it will take, we introduce a 
new stochastic variable $\lambda=\tsigma M$. Then \eqsand{eq:M2}{eq:sigma2} 
become, for $\xi|\lambda|>M^2$, 
\bea
\label{eq:Mlambda}
\dd_\tt M &=& \lambda,\\
\dd_\tt \lambda &=& \frac{\lambda^2 - \sqrt{\xi|\lambda|}\,\lambda}{M}
+ \frac{\sqrt{2}\lt(\xi|\lambda|\rt)^{3/4}}{\sqrt{M}}\,\chi(\tt).  
\label{eq:lambda}
\eea
The Fokker--Planck equation for the joint PDF of $M$ and $\lambda$ is 
obtained in the same fashion as it was done in \apref{ap:zeroD} for the PDF of $B$ and $\sigma$. 
The result is 
\bea
\nonumber
M\bl(\dd_\tt P + \lambda\dd_M P\br) &=& 
\xi^{3/2}\dd_\lambda|\lambda|^{3/4}\dd_\lambda|\lambda|^{3/4} P\\
&&+\dd_\lambda\lt(\sqrt{\xi|\lambda|}\,\lambda - \lambda^2\rt) P \equiv \LL P.   
\eea
This equation has a self-similar solution
\beq
P(\tt,M,\lambda) = \frac{1}{\tt}\,f\lt(m,\lambda\rt),\quad m = \frac{M}{\tt},
\eeq
where $f(m,\lambda)$ satisfies
\beq
-m\dd_m\lt(m - \lambda\rt) f  = \LL f. 
\label{eq:f}
\eeq
Note that the PDF of $\lambda$ is, therefore, stationary, while the PDF of $M$ is self-similar, 
$P(\tt,M) = (1/\tt)\int\rmd \lambda f(M/\tt,\lambda)$. 

The shape of these PDFs is not hard to work out. 
\Eqref{eq:f} has the following simple solutions for $m\gg\lambda$ and $m\ll\lambda$: 
\bea
\label{eq:fm}
m\gg\lambda: && f = \frac{1}{m}\,f_0(\lambda),\\
m\ll\lambda: && f = f_0(\lambda),
\eea 
where $f_0$ satisfies 
\beq
\LL f_0 = 0. 
\label{eq:f0}
\eeq
This solution suggests that if we integrate out the $\lambda$ dependence, 
we should get a PDF of $M$ that is constant at $M\ll\tt$ and 
has an $M^{-1}$ tail at $M\gg\tt$ --- as is indeed confirmed by the 
numerical simulations (see the right panel of \figref{fig:M2_pdfs}). 
Note that, technically speaking, this PDF is not normalizable, 
but, as we explained in \secref{sec:M2stoch}, implicitly we assume a cutoff 
at $M\sim1$. 

The PDF of $\lambda$, which is 
the solution of \eqref{eq:f0}, is obtained via direct integration, with a 
stipulation that $f_0\to0$ as $\lambda\to\infty$.  
Introducing a new variable $x = |\lambda|/\xi$, \eqref{eq:f0} becomes
\beq
\dd_x x^{3/4}\lt(\dd_x + 1 \mp \sqrt{x}\rt)x^{3/4}f_0^\pm = 0, 
\eeq
where the upper (lower) sign is for $\lambda>0$ ($\lambda<0$). 
At the mirror threshold (the upper sign), 
\beq
f_0^+ = \frac{\const}{x^{3/4}}\,e^{\frac{2}{3}x^{3/2}-x}
\int_x^\infty\frac{\rmd y}{y^{3/4}}\,e^{-\frac{2}{3}y^{3/2}+y};
\label{eq:fplus}
\eeq
at the firehose threshold (the lower sign),
\beq
f_0^- = \frac{\const}{x^{3/4}}\,e^{-\frac{2}{3}x^{3/2}-x}
\lt[\int_0^x\frac{\rmd y}{y^{3/4}}\,e^{\frac{2}{3}y^{3/2}+y} + c_0\rt].
\label{eq:fminus}
\eeq
The integration constants (the prefactors and $c_0$)
are fixed by normalization and matching $f_0^\pm$ to the 
behaviour at $x\lesssim M^2/\xi^2$, where the rate of strain 
is within the stability interval and so \eqref{eq:sigma2_stable} 
must be used --- we will not go into these complications here. 
The distributions \exref{eq:fplus} and \exref{eq:fminus} 
are power laws both at small and large $x$: 
\bea
x\ll1:&& f_0^+ \sim \frac{1}{x^{3/4}},\quad
f_0^- \sim \frac{c_0}{x^{3/4}} + \frac{4}{\sqrt{x}},\\ 
x\gg1:&& f_0^+ \sim f_0^- \sim \frac{1}{x^2}. 
\eea 
\Figref{fig:M2_pdfs} (left panel) shows $f_0^+$ found in our 
numerical simulations, with both power laws in reasonable 
agreement with theory. 

\bibliographystyle{mn2e}
\bibliography{ms_MNRAS}

\label{lastpage}

\end{document}